\documentclass[12pt]{article}

\usepackage{latexsym,amsmath,amssymb,pstricks,wasysym,
stmaryrd,enumerate,epsfig}

\newcommand{\be}{\begin{equation}}
\newcommand{\ee}{\end{equation}}
\newcommand{\beu}{\begin{equation*}}
\newcommand{\eeu}{\end{equation*}}
\newcommand{\bea}{\begin{eqnarray}}
\newcommand{\eea}{\end{eqnarray}}
\newcommand{\beaa}{\begin{eqnarray*}}
\newcommand{\eeaa}{\end{eqnarray*}}
\newcommand{\bmx}{\begin{pmatrix}}
\newcommand{\emx}{\end{pmatrix}}

\newcommand{\del}{\partial}
\newcommand{\g}{{\frak g}}
\newcommand{\h}{{\frak h}}
\newcommand{\m}{{\frak m}}

\newcommand{\am}{{\alpha}}
\newcommand{\bm}{{\beta}}
\newcommand{\gm}{{\gamma}}
\newcommand{\A}{{\cal A}}

\newcommand{\X}{I}
\newcommand{\I}{\Phi}

\newcommand{\ah}{{\hat \alpha}}
\newcommand{\bh}{{\hat \beta}}
\newcommand{\gh}{{\hat \gamma}}
\newcommand{\deh}{{\hat \delta}}

\newcommand{\roh}{{\hat \rho}}
\newcommand{\kh}{{\hat \kappa}}
\newcommand{\J}{{\cal J}}
\newcommand{\D}{{\cal D}}
\newcommand{\G}{{\cal G}}
\newcommand{\K}{{\cal K}}
\newcommand{\C}{{\cal C}}
\newcommand{\HH}{{\cal H}}
\newcommand{\dxy}{\delta(x{-}y)}
\newcommand{\dpxy}{\delta'(x{-}y)}
\newcommand{\half}{{\textstyle{1\over2}}}

\newcommand{\nn}{\nonumber}

\newcommand{\de}{\delta}
\newcommand{\eps}{\epsilon}
\newcommand{\tr}{{\rm Tr}}

\newcommand{\rank}{{\rm rank}}
\newcommand{\Pf}{{\rm Pf}}

\newcommand{\hc}{{\dagger}}

\advance\textheight by \topskip
\begin{document}
\baselineskip 17pt
\parindent 8pt
\parskip 8pt

\begin{flushright}
DAMTP-2004-90
\break
hep-th/0501090\\[3mm]
\end{flushright}

\begin{center}
{\Large {\bf Higher-spin conserved currents in supersymmetric
sigma models on symmetric spaces}}\\
\vspace{0.8cm} 
{\large J.M. Evans${}^a$,
C.A.S. Young${}^{a,b}$
}
\\
\vskip 5pt
{\em ${}^a$DAMTP, Centre for Mathematical Sciences, University of
Cambridge,\\ Wilberforce Road, Cambridge CB3 0WA, UK}
\\
\vskip 5pt
{\em ${}^b$Department of Mathematics, University of York,\\
Heslington Lane, York YO10 5DD, UK}
\\
\vskip 5pt
{\small E-mail: {\tt J.M.Evans@damtp.cam.ac.uk}, 
{\tt C.A.S.Young@damtp.cam.ac.uk}}
\\
\end{center}
\vspace{0.5cm}

\vskip 0.15in
 \centerline{\small\bf ABSTRACT}
\centerline{
\parbox[t]{5in}{\small
Local higher-spin conserved currents are constructed in the
supersymmetric sigma models with target manifolds symmetric spaces 
$G/H$. One class of currents is based on generators of the de Rham 
cohomology ring of $G/H$; a second class of currents are higher-spin 
generalizations of the (super)energy-momentum tensor. A comprehensive 
analysis of the invariant tensors required to construct these currents 
is given from two complimentary points of view, and sets of primitive 
currents are identified from which all others can be constructed as 
differential polynomials. The Poisson bracket algebra of the top
component charges of the primitive currents is calculated. It is shown
that one can choose the primitive currents so that the bosonic charges 
all Poisson-commute, while the fermionic charges obey an algebra which 
is a form of higher-spin generalization of supersymmetry. Brief 
comments are made on some implications for the quantized theories.
}}

\vspace{20pt}
\section{Introduction}

Non-linear sigma models in $1{+}1$ dimensions provide a fascinating
set of highly non-trivial quantum field theories which have been 
intensively studied for many years \cite{Zinn-Justin}-\cite{AF}. 
Of particular note are those whose target manifolds are symmetric
spaces, since these are known to be
integrable classically \cite{nonlocal,EFlocal} and, in some cases,
quantum-mechanically
\cite{Luscher,Fendley,Babichenko,EKY,GW,EKMY}. 
As well as being of much intrinsic interest
and displaying rich mathematical structures,  
these, or closely related models, are currently subjects of active 
study in fields as diverse as string theory
\cite{Mandal,Bena,Dolan,Tseytlin} and 
condensed matter physics (see
e.g.~\cite{Zinn-Justin,Fendley,Fendley:2002} and
references therein.

It was shown in \cite{PCM,Evans01} that bosonic principal chiral
models (PCMs), i.e.~sigma models with target spaces Lie groups, contain 
classical, commuting charges with spins equal to the exponents of the 
Lie algebra modulo its Coxeter number.
The corresponding currents are local functions of the
underlying fields and are constructed using totally symmetric
invariant tensors on the Lie algebra.
Some of these conserved charges are known to survive in the quantum
theory \cite{GW} and it is reasonable to conjecture that they all do, 
since their existence provides an explanation of common 
features shared by S-matrices for PCMs and affine Toda theories
(which are, physically, quite different models see e.g.~\cite{Corrigan}).
In so doing, they also fit together with known properties of non-local 
charges in a highly non-trivial, and rather intriguing, fashion.
Analogous results have been established for the wider class of 
bosonic sigma models whose target manifolds are symmetric spaces.
Not all such models are quantum integrable, and exactly when this
happens has recently been clarified, in terms of both 
local and non-local quantum charges \cite{EKY,EKMY}.

Supersymmetric extensions of the PCMs 
were considered in \cite{SPCM} (which also contains extensive
references to the earlier literature). 
Local, higher-spin conserved currents were shown to appear in two families:
in terms of an underlying fermionic superfield 
current $J^a$ taking values in the Lie algebra (more detailed
conventions follow) these currents take the form 
\be
\Omega_{a_1 a_2 \dots a_{p}} J^{a_1} J^{a_2} \ldots J^{a_n} 
\qquad {\rm and} \qquad
\Lambda_{a_1\ldots a_{q-1} a_{q}} J^{a_1} \ldots J^{a_{q-1}} D J^{a_q}
\ee
(with $D$ a superspace derivative).
The tensors $\Omega$ are invariant and totally antisymmetric and can 
be identified with cohomology generators of the Lie group.
The tensors $\Lambda$ are less familiar mathematically, but the 
first member of this sequence is nothing but the super-energy-momentum 
tensor. 
Each of these families gives rise to bosonic conserved charges 
whose spins are, once again, the exponents of the Lie algebra, 
but now with no repetition modulo the Coxeter number.
This strong similarity with the bosonic PCMs is actually 
rather surprising, because the 
$\Omega$ and $\Lambda$ currents above are \emph{not}
merely super-extensions of the currents in the bosonic theory, rather 
they are intrinsic to the supersymmetric PCMs, in the sense that they
vanish when the fermions are set to zero.
It was also shown in \cite{SPCM} that 
the currents can be chosen so that the resulting 
bosonic conserved charges all Poisson-commute.

In this paper we will extend many of the results above
and set them in a more general context by constructing and 
studying local conserved quantities in supersymmetric sigma models 
with target manifolds a compact symmetric space $G/H$. 
We will concentrate on the cases in which 
$G$ and $H$ are classical groups, namely \cite{Helgason}:
\bea
&SO(p{+}q)/SO(p){\times}SO(q) \ , \quad 
SU(p{+}q)/S(U(p){\times}U(q)) \ , \quad 
Sp(p{+}q)/Sp(p){\times}Sp(q) \ , & \nn \\  
&SU(n)/SO(n) \ , \quad
SU(2n)/Sp(n) \ , \quad 
SO(2n)/U(n) \ , \quad 
Sp(n)/U(n) \, &
\eea
although some of our methods could be used to 
treat the remaining examples, involving exceptional groups, too.
The first task is to construct generalizations of the 
$\Omega$ and $\Lambda$ currents above, which involves some 
interesting mathematical questions---in particular, how the cohomology 
of $G/H$ enters. After introducing the models and taking care of other 
preliminary matters in section 2, we carry out the construction 
in sections 3 and 4. We summarize our results in section 5,
identifying a set of `primitive' currents from which all others 
can be found.
A corollary of our construction is a rather concrete description
of generators for the cohomology ring of $G/H$, which may be useful 
in other contexts.
 
In the remainder of the paper we consider the current 
algebra of these conserved quantities, starting by setting up the
canonical formalism for the models in section 6.
Although the results we obtain are all classical, a strong
motivation for our work is the possibility of eventually extending 
them to the quantum level (all the supersymmetric models on symmetric
spaces are known to be quantum integrable \cite{AF})
and we make a number of comments on the likely implications 
for the quantum theory throughout the course of the paper.
In particular, the `top component' charges of the conserved currents 
are the ones most likely to survive in the quantum theory,
for reasons explained in section 5. 
We investigate their classical Poisson bracket algebra in section 7, 
and prove the existence of mutually commuting sets
of bosonic charges, thereby generalizing the work in \cite{SPCM}.
A completely new feature of certain symmetric space models is 
the th existence of top component fermionic charges
whose algebra closes in a kind of higher-spin generalization 
of supersymmetry. We conclude with some suggestions for future
research in section 8.

\section{The $G/H$ sigma model in superspace}

\subsection{Setting up the theory}

To formulate the theory in a manifestly supersymmetric fashion, we will
work in superspace 
with coordinates $\left\{x^{\pm}, \theta^{\pm} \right\}$, where 
$ x^{\pm}=\half ( x^0 \pm x^1) $
are the usual light-cone coordinates on two dimensional Minkowski
space, and the additional fermionic coordinates $\theta^{\pm}$ are 
real Grassmann numbers. The supersymmetry generators are
\be Q_{\pm} = \del_{\theta^{\pm}} + i \theta^{\pm} \del_{\pm}\ ,\ee and the
supercovariant derivatives are \be D_{\pm} = \del_{\theta^{\pm}} - i
\theta^{\pm} \del_{\pm} \ .\ee 
These obey
\be Q_\pm^2=i\del_\pm \ , \qquad  D_\pm^2=-i\del_\pm \ ,
\label{SUSYalgebra}\ee
with all other anti-commutators vanishing. Note that a Lorentz
boost of rapidity $\lambda$ acts by
$x^\pm \mapsto e^{\pm \lambda} x^\pm$ and 
$\theta^\pm \mapsto e^{\pm \lambda/2} \theta^\pm$ on superspace
coordinates, but by $\del_\pm \mapsto e^{\mp \lambda} \del_\pm$ and 
$D_\pm \mapsto e^{\mp \lambda/2} D_\pm$ on derivatives. In general, the 
\emph{spin}, or \emph{Lorentz weight}, of any quantity can be read-off 
by counting $\pm$ indices appropriately.

We begin by recalling the superspace formulation of the supersymmetric 
principal chiral model (SPCM) with target space a Lie group.
Let $\G(x,\theta)$ be a superfield 
taking values in a compact Lie group $G$, with Lie algebra $\g$, 
from which we define a fermionic superfield current
\be 
\J_{\pm} = \G^{-1} D_{\pm} \G \ , \qquad {\rm with} \qquad 
i \J \in \g
\label{Jdef}\ee 
(the factor of $i$ here may seem strange but is typical of 
the sorts of reality conditions that arise
when there are underlying Grassmann quantities).
The superspace lagrangian for the SPCM on $G$ is then
\be \mathcal L_G=-{\textstyle \frac{1}{2}} \, {\rm Tr} 
\left( \J_+ \J_- \right) \label{PCMlag}\ee
which is invariant under a global symmetry $G_L \times G_R$
acting on the superfield $\G$ by left and right multiplication
(for a fuller discussion, see e.g.~\cite{SPCM}).

Now consider a subgroup $H$ of $G$; let $\h$ be its Lie algebra 
and $\m$ the orthogonal complement of this in $\g$. 
The condition for $G/H$ to be a symmetric space is 
\be \g=\h + \m  \qquad {\rm where} \qquad 
[\h,\h]\subset\h \ , \quad [\h,\m]\subset\m \ , \quad[\m,\m]\subset\h
\ . 
\label{gradeg}\ee
To fix notation: we will often make use of an orthonormal basis
$\{t_a\}$ of generators of $\g$ (anti-hermitian matrices in the
defining representation) obeying 
\be 
[t_a,t_b] \ = \ f_{abc} \, t_c \ , \qquad 
{\rm Tr}( t_a t_b) = -\delta_{ab} \ .
\label{basis}\ee
We may choose $\{t_a\}$ to be the disjoint 
union of a basis $\{t_\ah\}$ of $\h$ and a basis $\{t_\am\}$ of $\m$. 
Any quantity in the Lie algebra $X \in \g$
can be written 
\be X \ = \ X^a t_a \ = \ X^{\ah} t_{\ah} + X^{\am} t_{\am} \ , \ee
thereby decomposing it into parts belonging to $\h$ and $\m$.
(There is no distinction between 
upper or lower Lie algebra indices.)
The symmetric space conditions 
(\ref{gradeg}) imply that 
\be 
f_{\ah \bh \gamma} = f_{\am \beta \gamma \vphantom{\bh}} = 0
\label{gradef} \ee
so that $f_{\ah \bh \gh}$ and $f_{\am \bm \gh}$ are 
the only non-vanishing structure constants, up to 
permutations of indices. From the Jacobi identity, these satisfy
\be
f^{\ah}{}_{\am \gm} f^{\bh}{}_{\gm \bm}  - 
f^{\bh}{}_{\am \gm} f^{\ah}{}_{\gm \bm}  = f^{\ah \bh \gh} f^{\gh}{}_{\am \bm}
\ , \qquad 
f^{\ah}{}_{ \am [\bm} f^{\ah}{}_{\gm \delta ]}  = 0 \ .
\ee

To define the supersymmetric $G/H$ sigma model, let $\G(x , \theta)$ be a
superfield taking values in $G$, as before, with $\J_\pm$ defined by
(\ref{Jdef}), and let 
\be \J_\pm = \K_\pm + \A_\pm \qquad {\rm where} 
\qquad i\K \in \m \, , \quad i \A \in \h 
\label{JKA}\ee
The superspace lagrangian for the $G/H$ model is then
\be \mathcal L_{G/H} =
-{\textstyle \frac{1}{2}} \, {\rm Tr} \left( \K_+ \K_- \right) \ .
\label{G/Hlag}\ee  
This is invariant under a global $G$ symmetry
$\G \mapsto U \G$ with $U \in G$ and a local $H$ gauge symmetry
$\G \mapsto \G \HH$ for any superfield $\HH( x ,\theta) \in H$.
It is useful to define a superspace derivative which is covariant
with respect to this gauge symmetry:
\bea
\D_\pm & = & D_\pm + \A_\pm \ , \\
\A_\pm & \mapsto & 
\HH^{-1} \A_\pm \HH + \HH^{-1} D_\pm \HH \ .
\label{Agauge} \eea
and to note that 
\bea
\K_\pm & = & \G^{-1} \D_\pm \G \, = \, \G^{-1} D_\pm \G - \A_\pm \ , \\
& \mapsto & 
\HH^{-1} \K_\pm \HH \label{Kgauge} \ .
\eea
The lagrangian is clearly gauge-invariant, with the physical 
degrees of freedom confined to the 
coset space $G/H$, as desired.\footnote{The superfields $\A_\pm$ could, 
alternatively, be introduced as independent variables, but they are
then non-dynamical, with algebraic equations of motion given by 
(\ref{JKA}).}

The superspace equations of motion following from the 
lagrangian can be written
\be
\D_+ \K_- - \D_- \K_+ \ = \ 0
\ee
and in addition we have, identically,
\be D_+ \J_- + D_- \J_+ + [ \J_+, \J_- ] \ = \ 0 \ .\ee
Here, and throughout the paper, Lie algebra brackets 
of bosonic or fermionic quantities 
will always be understood to be graded appropriately (so, for
example, $F^2 = \frac{1}{2} [ F , F]$ for a fermionic Lie
algebra-valued quantity).
Using the symmetric space property (\ref{gradeg}) 
we find 
\be  \D_\mp \K_\pm \ = \ D_\mp \K_\pm \, + \, [ \A_\mp , \K_\pm ] \ =\ 0
\label{laxeqn} \ee
and
\be [ \K_+, \K_- ] \ = \ D_+ \A_- + D_- \A_+ + [\A_+, \A_- ] \ = \ 0 \ .
\label{curvature}\ee
These equations are of a rather special (Lax) form 
which implies the classical integrability of
the model. One way to establish this is to construct 
\emph{non-local} conserved quantities, as described in
\cite{susynonlocal,SPCM}.
Our aim in this paper is 
to investigate other exotic conserved quantities which, by contrast, are
\emph{local} in the sigma model fields.

\subsection{Local conserved currents; $\Omega$ and $\Lambda$ tensors}

We will be concerned with superfield currents 
$\C$ which obey conservation equations of 
\emph{super holomorphic} type, meaning 
\be 
D_- \C = 0 \qquad {\rm or} \qquad \C 
= r + \theta^+ s \qquad
{\rm with} \qquad \del_- r = \del_- s = 0 \ .
\label{shol}\ee
Note that such a current $\C$ has a component expansion 
of simplified form. 
The standard conserved charges 
arising from the component currents $r$ and $s$ are
\be R = \int dx \, r \qquad \text{and} \qquad S = \int dx \, s \ , \ee 
one of which is bosonic and the other fermionic (which is which
depends on the grading of $\C$). 
We shall refer to $s$ and $S$ as the \emph{top component} current and 
charge, and to $r$ and $R$ as the \emph{bottom component} current and 
charge. We could, of course, equally well consider 
\emph{anti}-holomorphic currents, which are annihilated by $D_+$.

Given a set of superholomorphic quantities, we can take  
arbitrary polynomials in members of this set and their 
$D_+$ derivatives to obtain new superholomorphic 
expressions.
Note, however, that if $\C$ obeys (\ref{shol}) then 
\be
D_+ \C = s - i \theta^+ \del_+ r \ .
\ee
Although this expression is certainly superholomorphic, its top component
charge vanishes, while its bottom component charge 
is just $S$ again. Differential polynomials which are \emph{total} 
derivatives can therefore be disregarded because they yield 
nothing new. More generally, we shall refer to a superholomorphic 
quantity as \emph{composite} if it can be written as a non-trivial
differential polynomial in other super-holomorphic quantities, and
as \emph{primitive} if it cannot.
It is then natural to seek a set of primitive \emph{generators},
in terms of which all other superholomorphic currents 
can be expressed as differential polynomials.

To construct superholomorphic quantities $\C$ from the 
gauge-covariant currents $\K_+$, requires 
knowledge of $H$-invariant tensors on $\m$.
By definition, such a tensor $T$ obeys
\be T (X,Y,\dots,Z) \ = \
T (hXh^{-1}, hYh^{-1}, \dots, h Zh^{-1})
\label{Tinv1}\ee
for all $X,Y,\dots,Z\in\m$ and all $h\in H$;
or equivalently
\be T( \, [X,W], \, Y, \ldots, \, Z \, ) \ + \
T(\, X, \, [Y,W], \, \ldots, \, Z) \ + \
\dots+T (\, X, \, Y, \, \ldots, \, [Z,W] \, ) \ = \ 0 
\label{Tinv2}\ee
for all $X,Y,\ldots,Z\in\m$ and $W \in \h$.
With respect to a basis as in (\ref{basis}) and (\ref{gradef}),
we have components 
\be
T( X , Y , \ldots , Z) = 
T_{\am_1 \am_2 \ldots \am_p} X^{\am_1} Y^{\am_2} \ldots Z^{\am_p}
\qquad {\rm or} \qquad 
T_{\am_1 \ldots \am_p} = T (t^{\am_1} , \ldots , t^{\am_p}) \ ,
\label{Tcomp}\ee
and the condition for invariance is 
\be
T_{\bm \am_2 \ldots \am_p} f^{\bm}{}_{\am_1 \hat \gamma} 
+ T_{\am_1 \bm  \ldots \am_p} f^{\bm}{}_{\am_2 \hat \gamma} 
+ \ldots 
+ 
T_{\am_1 \am_2 \ldots \bm } f^{\bm}{}_{\am_p \hat \gamma} 
= 0 \ .
\label{Tinv3}\ee

Consider first the possibility of currents multilinear in 
$\K_+$.
Let 
\be \C_\Omega \ = \ \Omega \left (\K_+, \K_+ , \ldots, \K_+\right) \ = \ 
 \Omega _ {\am_1 \am_2 \dots \am_p} \, \K_+^{\am_1} \K_+^{\am_2} 
\ldots \K_+^{\am_p} \ , 
\label{Ocurr}\ee
where $\Omega$ is an invariant tensor on $\m$ which we may take to 
be totally antisymmetric (since $\K_+$ is fermionic). 
For future reference we note that the invariance of $\Omega$ 
can then be written
\be
\Omega_{\bm [\am_2 \ldots \am_p} f^{\bm}{}_{\am_1 ] \hat \gamma} = 0 \ .
\ee
Invariance of $\Omega$ implies, immediately, that $\C_\Omega$ is 
superholomorphic from the equations of motion (\ref{laxeqn}):
because $\C_\Omega$ is gauge-invariant, the derivatives $D_-$ and
$\D_-$ agree on it, and so 
\be D_-\C_\Omega \ = \ \D_-\C_\Omega \ \propto \ \Omega _{\am_1 \dots \am_p} \,
\K_+^{\am_1} \ldots \K_+^{\am_{p-1}} \, \D_-\K_+^{\am_p} \ = \ 0 \ .
\ee
The resulting top and bottom component currents have spins 
$(p{-}1)/2$ and $(p{-}2)/2$, respectively.

A second class of superconformal currents arise as 
higher-spin generalizations of the super-energy-momentum tensor.
The general formula is
\be \C_\Lambda \ = \ \Lambda \left( \K_+, \dots ,\K_+, ; \D_+\K_+ \right) 
\ = \ \Lambda_{\am_1 \dots \am_q ; \bm} \,  
\K_+^{\am_1} \ldots \K_+^{\am_q} \, \D_+ \K_+^\bm
\label{Lcurr} \ee
where $\Lambda$ is an $H$-invariant tensor on $\m$ 
which is antisymmetric on all its indices except the last (which is 
separated from the others by a semi-colon for this reason).
The super-energy-momentum tensor $\K_+^\am D_+ \K_+^\am$ 
is obtained for $\Lambda_{\alpha
\beta}= \delta_{\alpha \beta}$.
Notice that $\Lambda$ must never be \emph{totally} antisymmetric,
or else the expression above will be a total derivative.
Now $H$-invariance alone is not enough to ensure that 
$\C_\Lambda$ is superholomorphic:
\be D_- \C_\Lambda \ = \ \D_- \C_\Lambda \ = \ 
\Lambda_{\am_1 \dots \am_q; \, \bm} \, 
\K_+^{\am_1} \ldots \K_+^{\am_q} \, \D_- (\D_+\K_+^\bm)\ee
using (\ref{laxeqn}), but then 
\be \D_- (\D_+ \K_+) 
\ = \ -[ \, [ \K_-,\K_+], \, \K_+ \, ]\ee 
using (\ref{curvature}). So  
$\C_\Lambda$ is superholomorphic iff 
\be \Lambda_{ [ \am_1 \dots \am_q;}{}^{\bm} 
f^{\bm \gh}{}_{\de} f^{ \gh \eps }{}_{\lambda ] } \ = \ 0 \ .
\label{lambdaprop}\ee
When this holds, the resulting top and bottom component 
currents have spins $(q{+}1)/2$ and $q/2$, respectively.

Our task is to construct
tensors $\Omega$ and $\Lambda$ with these properties
for each classical symmetric space $G/H$.
We will do this in sections 3 and 4, using 
two complimentary approaches. 
The first approach is rather general and could be applied to
examples involving exceptional groups too (though we do not
consider such cases in any detail in this paper). 
It involves starting from symmetric 
$G$-invariant tensors on $\g$, or $H$-invariant tensors on $\h$, 
and using these to build $H$-invariant tensors on $\m$ in a systematic
way. The second approach is more case-specific and involves
writing down $H$-invariants on $\m$ directly, using 
knowledge of the particular representation of $H$ on $\m$ for each 
classical symmetric space. 

The combination of these approaches will 
allow us to identify generating sets of primitive currents,
as defined above. This will depend, in part, on understanding 
whether the invariant tensors used to define the currents are
\emph{primitive} as tensors, meaning that they 
cannot be written as tensor products 
(appropriately symmetrized or antisymmetrized) 
of invariants of lower degree, or whether they are
\emph{compound}, meaning that they \emph{can} be decomposed 
in such a fashion. 
The relationship between the notions of primitive
tensors and primitive superholomorphic 
currents is quite intricate, however, as we shall see.

\section{General approach}

\subsection{Lie groups---review}

We begin by recalling some details concerning invariants for 
Lie groups and Lie algebras. This will be essential for understanding
the generalization to symmetric spaces and many of the details will
also be needed at other points throughout the paper.

Note that $G$-invariant tensors on $\g$ are defined by the equations 
(\ref{Tinv1}) to (\ref{Tinv3}) given earlier, with $\m$ replaced by
$\g$ and $H$ replaced by $G$.
For a simple Lie algebra $\g$ there are 
exactly ${\rm rank} (\g)$ independent primitive symmetric 
$G$-invariant tensors $d$ (see e.g.~\cite{Azcarraga})
The choice of these primitive invariants is certainly not unique,
because we always have the freedom to add on products of invariants of
lower degrees,
but once we have chosen a particular set, then \emph{any} other
symmetric invariant can be expressed in terms of them in a unique way.
Furthermore, the degrees $p$ of the tensors $d$ are the same for each
primitive set; writing $p=s+1$, the integers $s$ are
the \emph{exponents} of $\g$, and for the classical Lie algebras we have  
\bea 
\frak{a}_{n} = \frak{su} (n{+}1) \ \ & : & s = 1, \, 2, \, \ldots, \,n
\nonumber \\ 
\frak{b}_n = \frak{so} (2n{+}1) \ \ & : & s = 1, \, 3, \, \ldots , \, 2n{-}1
\nonumber \\ 
\frak{c}_n = \frak{sp} (n) \ \ & : & s = 1, \, 3, \, \ldots , \, 2n{-}1
\nonumber \\ 
\frak{d}_n = \frak{so}(2n) \ \ & : & 
s = 1, \, 3, \, \ldots , \, 2n{-}3; \, n{-}1 
\label{exp}
\eea

A standard way to construct symmetric invariant tensors is as 
symmetrized traces in some representation; for the classical
groups or algebras we can use the defining representations:
\be
{\rm Tr} (X^p) = 
  d_{a_1 a_2 \ldots a_p} \, X^{a_1} X^{a_2} 
\ldots X^{a_p} \qquad {\rm or} \qquad
  d_{a_1 a_2 \ldots a_p} = {\rm Tr} \left( t_{( a_1} t_{a_2} 
\ldots t_{a_p )} \right).
\label{gtr}\ee
One can always choose a set of primitive invariants from amongst
these, with one exception. For $\g = \frak{so} (2n)$ there is an 
invariant of degree $n$, called the Pfaffian, which is not of this
form; it is defined by 
\be
\Pf ( X ) \ = \ d_{a_1 a_2 \ldots a_n} \, X^{a_1} X^{a_2} \ldots X^{a_n}
\ = \ \frac{1}{2^n n!} \,
\varepsilon_{i_1 j_1 i_2 j_2 \dots i_n j_{n}} \, 
X_{i_1 j_1} X_{i_2 j_2} \ldots 
X_{i_{n} j_{n}} .
\ee
(This is related to a trace in a spinor representation.)

The values of $s$ in (\ref{exp}) can be understood using the 
following general fact. Given any $m{\times}m$ complex matrix, 
$X$, the trace-powers 
\be 
{\rm Tr} X \ , \ \ 
{\rm Tr} X^2 \ , \ \ 
\ldots  \ , \ \
{\rm Tr} X^m \  
\label{trpow} \ee
are independent, in general, but traces of all higher powers 
can always be expressed in terms of them.
This follows from the identity
${\rm Det }(1 -\lambda X) = \exp {\rm Tr} \log (1 -\lambda X)$;
the left-hand side is a polynomial in $\lambda$ of degree $m$
but the right-hand side can be expanded in a power series
(for suitable $\lambda$)
and equating coefficients yields the desired relations.
When, in addition, $X$ has specific
properties by virtue of belonging to some Lie algebra,
then some of the traces in (\ref{trpow}) can vanish,
and it is easy to check the details and recover (\ref{exp}). 
Once again, the Pfaffian in $\frak{so}(2n)$  
is something of a special case, but 
${\rm Pf} (X)^2 = {\rm Det} X$, which can be expressed in terms of 
the trace-powers (\ref{trpow}) for $m \leq 2n$, and this relation 
then means that 
${\rm Tr} X^{2n}$ is compound, consistent with (\ref{exp}).

With these preparatory remarks in mind, we recall how to construct 
$\Omega$ and $\Lambda$ tensors for groups, i.e.~for the
SPCMs \cite{SPCM}.
Our formulation of the general $G/H$ model in
section 2.1 can, of course, be specialized to the $G$ SPCM,
with lagrangian (\ref{PCMlag}), 
by taking $H$ to be the trivial subgroup (we will 
comment below on another way of viewing groups as symmetric
spaces---see section 3.2).
Both the $\Omega$ and $\Lambda$ tensors for groups are 
defined using a primitive symmetric invariant tensor $d$ on $\g$.
If $d$ has degree $s{+}1$, the resulting top component
currents (or charges) have spins $s{+}1$ (or $s$),
with $s$ listed in (\ref{exp}).

To define $\Omega$ tensors, we take a 
primitive symmetric invariant $d$, fill all but one of 
the slots with a Lie bracket, 
and antisymmetrize; so in components 
\bea
  \Omega_{a_1\dots a_{2s+1}}&=&
\frac{1}{2^s}f^{b_1}{}_{[a_1 a_2} \ldots f^{b_s}{}_{a_{2s-1} a_{2s}} 
                               d_{a_{2s+1}] b_1 \dots b_s}\nonumber\\
 &=&\frac{1}{2^s}f^{b_1}{}_{a_1 [a_2} \ldots f^{b_s}{}_{a_{2s-1} a_{2s}]} 
                               d_{a_{2s+1} b_1 \dots b_s} \ , 
\label{Omegag}\eea
where the second line follows from the invariance of $d$.
We will often write $\Omega^{(d)}$ to indicate the 
underlying symmetric invariant $d$ from which $\Omega$ is built.
Note that if we attempt to build a totally antisymmetric invariant 
tensor by putting Lie brackets in 
\emph{all} of the slots, we get zero---because 
invariance of $d$ and the Jacobi identity imply 
\be
f^{b_1}{}_{[a_1 a_2} \ldots f^{b_{s+1}}{}_{a_{2s+1} a_{2s+2}]}  
                               d_{b_1 \dots b_{s+1}} \  = \ 0 \ .
			       \label{allslots}\ee
This will prove important later.
The tensors $\Omega^{(d)}$
provide a set of generators 
for the algebra of $G$-invariant forms on $\g$, and they 
can consequently be identified with the generators for the
de Rham cohomology ring of $G$ (see e.g.~\cite{Azcarraga}).

For the $\Lambda$ family, we start once again from a 
symmetric invariant $d$ and define  
\be
\Lambda_{a_1\dots a_{2s-1}; \, a_{2s}}=
\frac{1}{2^{s-1}}f^{b_1}{}_{[a_1 a_2} \ldots f^{b_{s-1}}{}_{a_{2s-3} a_{2s-2}} 
        d_{a_{2s-1}] a_{2s} b_1 \dots b_{s-1}}
\label{Lambdag}\\
\ee
For groups, the key identity (\ref{lambdaprop}) becomes 
\be 
\Lambda_{ [ a_1 \dots a_q;}{}^{b} 
f^{b c}{}_{d} f^{ c e }{}_{f ] } \ = \ 0
\ee
which can be derived from (\ref{allslots}). Once again, we will often
write $\Lambda^{(d)}$ to indicate the dependence on $d$.

\subsection{Symmetric spaces}

It is not immediately clear how one should generalize the 
definitions (\ref{Omegag}) and (\ref{Lambdag}) from groups to 
symmetric spaces.
One possibility is to start with tensors $\Omega$ and $\Lambda$ 
on $\g$ and simply restrict them to $\m$, which will certainly give
$H$-invariant results.
We must then determine when these restrictions are non-zero, or independent,
however. Furthermore, the work in \cite{SPCM} suggests that the association
of $\Omega$ and $\Lambda$ with the underlying $d$ tensor is 
best kept as clear as possible. We therefore proceed as follows.

Recall that we may define the symmetric space $G/H$ by means of
an automorphism $\sigma$ of $\g$, with $\sigma^2=1$, 
the subspaces $\h$ and $\m$ being the eigenspaces of $\sigma$ with 
eigenvalues $\pm 1$. 
Now consider how a given symmetric $G$-invariant tensor $d$ on $\g$ behaves 
when its entries are acted on by $\sigma$. There are two
possibilities:
\be 
d ( \sigma (X), \sigma (Y), \ldots , \sigma (Z) )
\ = \ \eta \, d ( X, Y , \ldots , Z ) \ , \qquad \eta = \pm 1 \ , 
\label{sigmad}\ee
for all $X, Y, \ldots , Z \in \g$.
The first possibility, $\eta = 1$, obviously holds whenever $\sigma$
is an inner automorphism, but 
if $\sigma$ is not inner then $\eta = -1$ may occur for certain
$d$ (the relevant automorphisms are complex conjugation for
$SU(n)$ or $E_6$, and reflection for $SO(2n)$ \cite{Helgason}).
It is an easy matter to determine which tensors have $\eta = \pm 1$ 
for each symmetric space
and it follows from the behaviour of $\h$ and $\m$ under $\sigma$
that the components of $d$ have the following properties:
\bea
\eta = +1 \quad & \Rightarrow & \quad 
d_{\hat \am_1 \hat \am_2 \ldots \hat \am_s \hat \am_{s+1}} \neq 0 \, , \quad
d_{\hat \am_1 \hat \am_2 \ldots \hat \am_s \am_{s+1} } = 0 \, , \quad
d_{\hat \am_1 \hat \am_2 \ldots \hat \am_{s-1} \am_s \am_{s+1}} \neq 0
\, , \ \ldots
\phantom{XXX} 
\label{grade1}\\
\eta = -1 \quad & \Rightarrow & \quad 
d_{\hat \am_1 \hat \am_2 \ldots \hat \am_s \hat \am_{s+1}} = 0 \, , \quad
d_{\hat \am_1 \hat \am_2 \ldots \hat \am_s \am_{s+1} } \neq 0 \, ,
\quad \ldots 
\label{grade2}
\eea
In these equations, we mean that the relevant components 
need not be identically zero by virtue of (\ref{sigmad}); 
some specific components may vanish.

Given any $d$ with $\eta = +1$, the properties (\ref{grade1})
imply that we can build a $\Lambda$ tensor in much the 
same way that we did for groups:
\be
 \Lambda^{(d)}_{\am_1 \dots \am_{2s-1}; \bm} 
= \frac{1}{2^{s-1}}f^{\bh_1}{}_{[ \am_1 \am_2} 
             \ldots f^{\bh_{s-1}}{}_{\am_{2s-3} \am_{2s-2}} 
d_{ \am_{2s-1} ] \bm \bh_1 \dots \bh_{s-1} } \ . 
\label{Lambda}\ee
This is certainly $H$-invariant and, by careful use of the 
invariance of $d$, one can check that 
it also satisfies the additional property (\ref{lambdaprop}).
These tensors are, indeed, just the restrictions of the 
$\Lambda$ tensors (\ref{Lambdag}) form $\g$ to $\m$, but 
the relevant properties (\ref{grade1}) of the underlying $d$ tensor are 
now transparent. Note also that with this choice the definition 
(\ref{Lcurr}) can be written 
\be
\C_\Lambda \ = \ d (\K^2_+ , \ldots , \K_+^2, \K_+ , \D_+ \K_+) \ .
\label{Ldcurr}\ee

Given a tensor $d$ with $\eta = -1$, we cannot construct a $\Lambda$ tensor
(the expression above vanishes). But we can, instead, construct
an $\Omega$ tensor:
\be\Omega^{(d)}_{\am_1\dots \am_{2s+1}}
=\frac{1}{2^{s}}f^{\bh_1}{}_{[\am_1 \am_2} \ldots
  f^{\bh_s}{}_{\am_{2s-1} \am_{2s}} 
d_{\am_{2s+1}] \bh_1 \dots \bh_s} \ .
\label{Omegad}\ee
With this choice, (\ref{Ocurr}) becomes:
\be
\C_\Omega \ = \ d (\K^2_+ , \ldots , \K_+^2, \K_+ )
\label{Odcurr}
\ee
Once again, this $\Omega$ tensor 
is the restriction to $\m$ of a $G$-invariant tensor
(\ref{Omegag}) on $\g$.
These are not the only $\Omega$ tensors 
however. 

The problem of finding a set of generators for the algebra of 
$H$-invariant antisymmetric tensors, or forms,
on $\m$ has been considered by mathematicians;
it corresponds to finding a set of generators for the de Rham 
cohomology ring of $G/H$ \cite{Borel:1953, Baum:1968, Milnor, Spivak:vol5} 
(see also \cite{D'Hoker:1995} for an account in the physics literature). 
We have just learnt that there is one class of antisymmetric tensors,
or cohomology representatives, which are based on symmetric 
$G$-invariant tensors 
$d$ on $\g$ which vanish when restricted $\h$. For Lie groups, a
complete set of cohomology generators arises in this way, and they are
all of odd degree. 
But for symmetric spaces there may be additional $\Omega$ tensors 
of even degree. These are of the form
\be
\Omega^{(e)}_{\am_1 \dots \am_{2 k}} \ = \ 
\frac{1}{2^{k}} f^{\bh_1}{}_{[ \, \am_1 \am_2} \ldots 
f^{\bh_k}{}_{\am_{2 k - 1} \am_{2 k \, {\vphantom{\bh}} ]}} \, 
e_{\bh_1 \dots \bh_k}
\label{Omegae}
\ee
where $e_{\bh_1 \dots \bh_k}$ is a symmetric 
$H$-invariant tensor on $\h$ which is \emph{not} the restriction
of a $G$-invariant tensor on $\g$.\footnote{ 
Regarded as invariant forms on $G/H$, 
the $\Omega^{(e)}$ correspond to characteristic classes.
Thinking of $G \rightarrow G/H$ as a principal $H$-bundle with a 
connection, the Chern-Weil homomorphism (see for example 
\cite{Milnor, Spivak:vol5})
is a map from the ring of invariant polynomials of the structure group 
to the de Rham cohomology of the base space; the definition
(\ref{Omegae}) is essentially this map.}
The general formula (\ref{Ocurr}) then reads
\be
\C_\Omega \ = \ e( \, \K^2_+ , \, \K^2_+ , \ldots , \, \K_+^2 \, ) \ .
\label{Oecurr}
\ee

To understand why $e$ must \emph{not} be the restriction of 
a $G$-invariant tensor $d$ on $\g$, note that (\ref{allslots}),
together with the symmetric space conditions (\ref{gradef}),
implies that
\be 
f^{\bh_1}{}_{[\am_1 \am_2} \ldots f^{\bh_{s+1}}{}_{\am_{2s+1} \am_{2s+2}]}  
                               d_{\bh_1 \dots \bh_{s+1}}  = 0
\label{allslotsonm}\ee
so that the expression in (\ref{Omegae}) then vanishes.
Moreover, this identity implies 
that $\Omega^{(e)}$ will also vanish
if $e$ is a symmetrized tensor product $d \cdot e'$ where
$d$ is the restriction of any $G$-invariant tensor on $\g$ 
(of strictly positive degree, of course).
We must therefore choose a maximal set of $e$ tensors
which are independent modulo such tensor products in order to obtain a 
full set of even-degree forms.

Let us now return briefly to Lie groups as special examples of 
symmetric spaces. For the formulation of the corresponding
sigma models, it is convenient to think of the SPCMs as 
having target manifolds $G/H$ with $H$ taken to be the trivial
subgroup. But we may instead regard $G = G_L{\times}G_R/G$,  
which has the advantage that the numerator, consisting of a 
direct product of two copies of $G$, is the isometry
group of the manifold (as it should be for a symmetric space 
$G/H$). If we apply our general construction of $\Omega$ and $\Lambda$ 
tensors to this case, we start with
independent symmetric invariants $d_L$ and $d_R$ on the Lie algebras
$\g_L$ and $\g_R$ for each factor.
The automorphism which defines the symmetric space simply exchanges
the $L$ and $R$ factors, so that the denominator $G$ is the
diagonal subgroup of $G_L \times G_R$.
But this means that we always have combinations 
$d_L \pm d_R$ on $\g_L \oplus \g_R$ which are even/odd under the 
automorphism. Our general construction therefore gives a nice 
additional insight into why there are \emph{both} $\Omega$ and $\Lambda$
tensors associated with \emph{each} primitive invariant $d$ for the case
of groups.

\subsection{Choices of tensors and currents}

To summarize, we have found that for each 
symmetric space $G/H$ there are $\Omega$ and $\Lambda$ tensors which arise 
from symmetric $d$ and $e$ tensors as follows:
\begin{itemize}
\item $d$, $G$-invariant on $\g$ and even under the automorphism $\sigma$ 
\hfill \break
$\longrightarrow$ $\Lambda^{(d)}$, tensor of even degree
\item $d$, $G$-invariant on $\g$ and odd under the automorphism $\sigma$
\hfill \break
$\longrightarrow$ $\Omega^{(d)}$, form of odd degree
\item $e$, $H$-invariant on $\h$ and not the restriction
of a $G$-invariant on $\g$ 
\hfill \break
$\longrightarrow$ $\Omega^{(e)}$, form of even degree
\end{itemize}

In order that 
$\C_\Omega$ or $\C_\Lambda$ be primitive as superholomorphic 
currents it is necessary that $d$ or $e$ 
be primitive as symmetric tensors, although this is not 
usually a sufficient condition, as we shall explain below.
It is also still important to understand how 
the construction works if the underlying 
symmetric tensors $d$ or $e$ are compound, or how the 
results are modified if they are primitive but
changed by the addition of compound terms
\bea
d_{a_1 \dots a_p} 
& \rightarrow & d_{a_1 \dots a_p} \ + \ 
d'_{(a_1 \dots a_j} d''_{a_{j+1} \dots a_p)} \label{dchange}\\
e_{\ah_1 \dots \ah_k} 
& \rightarrow & e_{\ah_1 \dots \ah_k} \ + \ 
e'_{(\ah_1 \dots \ah_j} e''_{\ah_{j+1} \dots \ah_k)} \label{echange}
\eea
The modification of $d$ must, of course, have 
the same behaviour under $\sigma$ in order to allow the construction 
of either a $\Lambda$ or an $\Omega$ tensor.

It is an immediate consequence of 
(\ref{allslotsonm}) that $\Omega^{(d)}$ 
vanishes unless $d$ is primitive 
(and $d$ must be odd under $\sigma$).
For the same 
reason, $\Omega^{(d)}$ and $\C_\Omega$ are 
independent of any change (\ref{dchange}).
The $\Omega^{(e)}$ tensors \emph{are} modified by (\ref{echange}),
but only by wedge products of forms of lower degree
\be
\Omega^{(e)} 
\ \ \rightarrow \ \
\Omega^{(e)} \ + \ \Omega^{(e')}\wedge\Omega^{(e'')}
\ .\ee
This in turn implies that $\C_\Omega$ changes by an expression
$\C_{\Omega'} \, \C_{\Omega''}$.

The tensors $\Lambda^{(d)}$, for which 
$d$ must be even under $\sigma$, behave in rather more 
complicated ways. 
If there exist $d'$ and $d''$ of the correct degrees 
which are each odd under $\sigma$, then the  
modification (\ref{dchange}) results 
in a change in $\Lambda^{(d)}$ proportional to 
\be
\Omega^{(d')}_{[\am_1 \dots \am_{2m-1}} 
\Omega^{(d'')}_{\am_{2m} \dots \am_{2p-3}] \bm}
\ + \ 
\Omega^{(d'')}_{[\am_1 \dots \am_{2n-1}} 
\Omega^{(d')}_{\am_{2n} \dots \am_{2p-3}] \bm}  \label{Lamchange}
\ee
(this is discussed in some detail in \cite{SPCM} for the case of
groups) and the current $\C_\Lambda$ acquires 
extra terms of the form $\C_{\Omega'} D_+ \C_{\Omega''}$.
If $d'$ and $d''$ are even under $\sigma$, however, then 
(\ref{grade1}) and (\ref{allslotsonm}) imply that 
$\Lambda^{(d)}$ is unaffected.
Furthermore, if the symmetrized product 
of \emph{three} or more tensors is added to $d$
then $\Lambda^{(d)}$ is always unchanged, again
by (\ref{allslotsonm}). Even if $d$ is primitive,
however, $\Lambda^{(d)}$ need not be: it may 
decompose into a product of the form $\Omega \ldots \Omega' \Lambda'$,
implying that the conserved current is composite.
We explain this in detail in the next section.

The observations above enable us to keep track 
of the effects of using different 
choices of primitive symmetric invariants $d$ and $e$
to define conserved currents, and there are two sets in particular 
that will prove useful.
The simplest possibilities for many purposes are the symmetrized
traces (\ref{gtr}), or trace-powers (\ref{trpow}), and we will deal
mainly with these when determining the pattern of primitive $\Omega$ and
$\Lambda$ tensors in the next section.
On the other hand, it was shown in \cite{PCM,Evans01} that for any Lie group 
$G$ there exists another choice 
of primitive symmetric tensors on $\g$ such that any pair 
$d_{a_1 a_2 \dots a_p}$ and $\tilde d_{b_1 b_2 \dots b_q}$, say, 
satisfy
\be 
d_{c(a_1 \dots a_{p-1}}\tilde d_{b_1 \dots b_{q-2} ) b_{q-1} c}
\ = \ 
d_{c(a_1 \dots a_{p-1}}\tilde d_{b_1 \dots b_{q-2}  b_{q-1} ) c}
\ .
\label{special} \ee
This property proved crucial to the construction of commuting sets of 
conserved charges in both the bosonic and supersymmetric PCMs \cite{PCM,SPCM}.
The same invariants will be equally important in our treatment of 
Poisson brackets in section 7.

\section{Case-by-case construction}

In this second approach we find conserved currents 
directly, for each classical symmetric space. We are able to do 
this because the representations of $H$ on $\m$ are very
familiar, involving defining representations or their tensor products,
so that invariants can be found comparatively easily.
We will be able to understand which of them are
independent and primitive by using  
a variation on the result following (\ref{trpow}).

Let $X$ and $Y$ be any $m{\times}m$ complex matrices; then 
\be 
{\rm Tr} Y \ , \ \ 
{\rm Tr} X Y \ , \ \ 
{\rm Tr} X^2 Y \ , \ \ 
\ldots  \ , \ \
{\rm Tr} X^{m-1} Y \  
\label{newtrpow} \ee
are in general independent, but all expressions ${\rm Tr} X^{r} Y$ 
for $r \geq m$ can be expressed in terms of the
quantities in (\ref{newtrpow}).
This follows from the corresponding result for (\ref{trpow}) by
considering $X + \lambda Y$ and expanding to first order in $\lambda$.
Note that the results for (\ref{trpow}) and (\ref{newtrpow}) also 
hold if $X$ and $Y$ are constructed from 
Grassmann quantities, provided $X$ is bosonic, or even graded.
The matrix $Y$ in (\ref{newtrpow}) is allowed to be 
fermionic, if desired, because in the proof indicated above
the parameter $\lambda$ can also be fermionic---no difficulties arise 
because we are expanding only to first order in $\lambda$
(but higher powers of fermionic matrices could not be treated 
in this way.)

\subsection{Complex Grassmannians}
The complex Grassmannians are
$SU(p+q)/S(U(p)\times U(q))$, with $p\leq q$, say. 
The current $\K_+$ and its derivative $\D_+ \K_+$ belong to $\m$ and
so take the block forms 
\be 
\K_+ = \begin{pmatrix}0 & K \\ -K^\hc & 0 \end{pmatrix}
\ , \qquad
\D_+ \K_+ = \begin{pmatrix} 0 & L \\ -L^\hc & 0
\end{pmatrix} \ , \label{block} \ee 
where $K$ (fermionic) and $L$ (bosonic) are
complex $p{\times}q$ matrices.
An element 
\be h=\begin{pmatrix} A & 0 \\ 0 & B \end{pmatrix} \ \ 
\in \ \ H \ , \label{hblock} \ee
where $H=S(U(p){\times}U(q))$, acts on these matrices 
according to
\be
K \mapsto A K B^\hc \ , \qquad L \mapsto A L B^\hc \ .
\label {haction}\ee
It is clear that to form $H$-invariants we must take
traces of products of matrices such as $K K^\hc$ and $K L^\hc$.

If we use just the matrices $K$, we can construct currents
\be
\C = \tr \left [ (K K^\hc)^r \right ] 
\label{KKcurr}\ee
and, considering (\ref{trpow}) with $X = KK^\hc$,
they are independent for
$1 \leq r \leq p$. These currents correspond to 
primitive forms $\Omega^{(e)}$ of even degree.
The invariants $e$ are trace-powers on the Lie algebra 
of $\frak{u}(p)$, but with no contribution from $\frak{u}(q)$,
which is why these $e$ tensors are not restrictions to $\h$ of
$G$-invariants on $\g$. Note also that one could take 
traces of powers of the $q{\times}q$ matrix $K^{\hc} K$ instead, 
and obtain the same currents (by cyclicity of the trace). But it is 
then no longer manifest that these invariants cease to be primitive 
beyond the $p$th power.

The currents based on $\Lambda^{(d)}$ tensors involve 
$K$'s and a single $L$, for example
\be
\C = 
\tr \left [ \, (K K^\hc)^{s-1} (K L^\hc + L K^\hc) \, \right ] \ .
\label{KLcurr}\ee
Note that if a relative sign is introduced 
in the last factor in the trace, then this expression becomes a total 
derivative, so we discount this possibility.
Considering (\ref{newtrpow}) with $X = K K^\hc$ and $Y = (KL^\hc + L
K^\hc)$, these currents are primitive for 
$1 \leq s \leq p$. 
For $s>p$, the trace will factorize into a sum of traces of lower
powers; in terms of tensors we have a compound $\Lambda$ factorizing into 
products of primitive tensors of the form 
$\Omega \dots \Omega \Lambda$.
To compare with our general construction of the last section,
all the invariants on $\g = \frak{su}(p{+}q)$ are unchanged 
under the automorphism $\sigma$ defining the symmetric space. Hence there 
are no tensors $\Omega^{(d)}$---no cohomology generators of odd
degree---for these spaces. The currents above 
correspond to tensors $\Lambda^{(d)}$ in which 
each $d$ is a trace-power on $\g$.
 
\subsection{Real and quaterionic Grassmannians}
The real Grassmannians $SO(p+q)/SO(p){\times}SO(q)$, 
with $p\leq q$, are of course similar in many ways to the complex
family above. We have the same block forms (\ref{block}) and 
(\ref{hblock}) but with $K$ and $L$ now $p{\times}q$ real  
matrices and $H = SO(p){\times}SO(q)$ acting as in (\ref{haction}).
However, new features arise because the matrix $KK^T$ is 
antisymmetric.

The currents 
\be
\C \,= \, \tr \left [ (K K^T)^r \right ]
\label{grasstr}\ee
are non-zero only if $r$ is an even integer 
(the trace of any odd power of an antisymmetric matrix vanishes)
and they are primitive when $r \leq p$, as before.
These are the currents based on $\Omega^{(e)}$ tensors 
with $e$ a trace-power invariant on $\frak{so}(p)$.
But there are additional currents, in some cases, which can be 
constructed using $\varepsilon$ tensors.

If $p$ or $q$ is even, we have 
\bea 
\C & = & \varepsilon_{i_1 i_2 \dots i_{p-1} i_{p}} (KK^T)_{i_1 i_2} \ldots 
(K K^T)_{i_{p-1} i_{p}} \ , \label{pf1} \\
{\rm or} \qquad 
\C & = & \varepsilon_{j_1 j_2 \dots j_{q-1} j_{q}} (K^T K)_{j_1 j_2} 
\ldots (K^T K)_{j_{q-1} j_{q}} \ , \label{pf2}
\eea
which correspond to forms $\Omega^{(e)}$ 
with $e$ the Pfaffian invariant on $\frak{so}(p)$ or 
$\frak{so}(q)$, respectively. 
These complete the set of primitive cohomology generators of even
degrees, but there is one more generator of odd degree which 
occurs iff both $p$ and $q$ are odd. The current in question is 
\bea
\C & = &  \varepsilon_{i_1 i_2 \dots i_{p-2} i_{p-1} i_{p}} \, 
\varepsilon_{j_1 j_2 \dots j_{q-2} j_{q-1} j_{q}} \nn \\
&& \quad \times \, (KK^T)_{i_1 i_2} \ldots (KK^T)_{i_{p-2} i_{p-1}} \,
(K^T K)_{j_1 j_2} \ldots (K^T K)_{j_{q-2} j_{q-1}} \, 
K_{i_p j_q} \ ,
\eea
which is based on the form $\Omega^{(d)}$ with $d$ the Pfaffian
on $\frak{so}(p{+}q)$. It is not difficult to see
that this $d$ tensor is odd under the automorphism
$\sigma$ defining the Grassmannian
iff $p$ and $q$ are both odd, consistent with our general construction in 
the last section.

Turning now to the currents involving derivatives, we have
\be
\C \ = \
\tr \left [ \, (K K^T)^{s-1} (K L^T) \right ] \ .
\label{grasskl}\ee
However, $KL^T + L K^T$ being symmetric implies that this
expression vanishes if $s$ is even. In addition, the non-vanishing
currents are primitive for $s \leq p$.
These currents are based on tensors $\Lambda^{(d)}$ with $d$ a 
trace-power on $\g = \frak{so}(p{+}q)$. The trace-type $d$ tensors are 
always invariant under $\sigma$, and the fact that $d$ must be the
trace of an even power for an orthogonal algebra fits precisely  
with the fact that the currents above are non-vanishing only when 
$s$ is odd. Finally, if $p$ and $q$ are both even we can form
one additional independent current 
\bea
\C & = &  
\varepsilon_{i_1 i_2 \dots i_{p-3} i_{p-2} i_{p-1} i_{p}} \, 
\varepsilon_{j_1 j_2 \dots j_{q-3} j_{q-2} j_{q-1} j_{q}} \nn \\
&& \quad \times
(KK^T)_{i_1 i_2} \ldots (KK^T)_{i_{p-2} i_{p-1}} \,
(K^T K)_{j_1 j_2} \ldots (K^T K)_{j_{q-3} j_{q-2}} \, 
K_{i_{p-1} j_{q-1}} L_{i_p j_q} \ . \label{pf3} \qquad 
\eea
This is based on $\Lambda^{(d)}$ with $d$ the Pfaffian on
$\frak{so}(p{+}q)$. We stated above that when $p$ and $q$ are 
both odd this $d$ tensor changes sign under $\sigma$ and so gives rise to 
an $\Omega$ tensor, but when $p$ and $q$ are both even $d$ is 
unchanged by $\sigma$ and so gives rise to a $\Lambda$ tensor instead.

The quaternionic Grassmannians $Sp(p{+}q)/Sp(p){\times}Sp(q)$ behave
similarly to the real cases, but without any of the complications arising
from Pfaffians. The block forms (\ref{block}) and (\ref{hblock}) 
hold with $p$ and $q$ replaced by $2p$ and $2q$, and with 
a symplectic reality condition 
$K^* = J^{\vphantom{1}}_p K J_q^{-1}$ where $J_p$ and $J_q$ are 
symplectic structures of size $2p{\times}2p$ and $2q{\times}2q$
respectively. The currents (\ref{KKcurr}) are non-vanishing 
for $r$ even, while the currents (\ref{KLcurr}) are non-vanishing 
for $s$ odd, and each set is primitive for $r, s \leq 2p$.
The cohomology generators $\Omega$ are all of even degree, based on 
$e$ tensors; the $d$ tensors on $\g = \frak{sp}(p{+}q)$ are unchanged 
by the automorphism defining the Grassmannian and so 
always give rise to $\Lambda$ type tensors.

\subsection{$SU(n)/SO(n)$ and $SU(2n)/Sp(n)$}
Consider first $SU(n)/SO(n)$, which is defined by 
taking the automorphism $\sigma$ of $\g = \frak{su}(n)$ to be complex
conjugation. This clearly implies that $\h = \frak{so}(n)$, while 
$\m$ consists of symmetric, traceless, imaginary $n{\times}n$ matrices.
Furthermore, $\K_+ \in \m$ transforms under $h \in SO(n)$ according to
$\K_+ \mapsto h \K_+ h^{-1}$. The currents 
\be \C = \tr \left [ \, (\K_+^2)^{s} \K_+ \, \right ] \ , \ee
are non-vanishing only for $s$ even (consider transposing the
matrices) and they are independent for $s+1 \leq n$.
Traces of even powers of $\K_+$ vanish by cyclicity.
Similar considerations imply that the currents
\be \C = \tr \left [ \, (\K_+^2)^{s-1} \K_+ \D_+ \K_+ \, \right ] 
\label{anothercurr}\ee 
are non-vanishing only when $s$ is odd and (\ref{newtrpow}) suggests that
they are independent for $s \leq n$, but there is a subtlety here
(see below) and the case $s=n$ can be ignored.

Comparing to section 3, the currents above 
arise from $d$ tensors on $\g = \frak{su}(n)$ corresponding 
to the trace-powers ${\rm Tr} X^{s+1}$.
These are inert under $\sigma$ (complex conjugation) 
when $s$ is odd, giving a tensor 
$\Lambda^{(d)}$, but they change sign 
when $s$ is even, giving a tensor $\Omega^{(d)}$. 
Note that $d$ must itself be primitive, which requires 
$s + 1 \leq n$ and that this is a slightly stronger condition than the one 
suggested by (\ref{newtrpow}) for the currents (\ref{anothercurr}).
The reason is that (\ref{newtrpow}) reveals whether $\Lambda$ can 
be decomposed into products $\Omega \ldots \Omega \Lambda$, 
but for $s=n$, $d$ is compound and $\Lambda^{(d)}$ 
decomposes as in (\ref{Lamchange}).

When $n$ is even, there is one final current 
\be
\C \ = \ \varepsilon_{i_1 i_2 \dots i_{n-1} i_n} \,  
(\K_+^2)_{i_1 i_2} \ldots (\K_+^2)_{i_{n-1} i_n} \ .  
\label{pf4} \ee
This corresponds to an even degree form $\Omega^{(e)}$,
where $e$ is the Pfaffian on $\h = \frak{so}(n)$.

Now consider the family $SU(2n)/Sp(n)$. This is very 
similar, because the automorphism $\sigma$ is defined by 
complex conjugation 
together with conjugation by a symplectic structure.
The currents and invariants are therefore just like those 
of $SU(2n)/SO(2n)$, but without the Pfaffian, for all $n > 2$.
(The case $n=2$ deviates slightly from the general pattern because
of some special properties of low-dimensional representations; 
but this case has already been dealt with above as one of the real
Grassmannians, since $SU(4)/Sp(2) = SO(6)/SO(5)$).

\subsection{$SO(2n)/U(n)$ and $Sp(n)/U(n)$}
Consider first the family $SO(2n)/U(n)$. 
The embedding of $\h = \frak{u}(n)$ in
$\g = \frak{so}(2n)$ is defined by taking $A + iB$,
where the real $n{\times}n$ matrices $A$ and $B$ are antisymmetric and
symmetric respectively, and mapping it to 
$\begin{pmatrix} A & B \\ -B & A 
\end{pmatrix}$. Similarly, elements of $\m$, which are of the form
${\textstyle{ \begin{pmatrix} C & D \\ D & - C 
\end{pmatrix} } }$, where the $n{\times}n$ matrices 
$C$ and $D$ are both real and antisymmetric, can be 
represented in the complex combination $C + i D$.
Thus the current $\K_+ \in \m$ can be 
identified with an $n{\times}n$, complex, antisymmetric matrix
$K$, and it can be checked that under 
$h \in U(n)$ this transforms $K \mapsto h K h^T$. 
In addition, $\D_+ \K_+$ is identified with a similar 
matrix $L$. 

Now (just as for the complex Grassmannians) one obvious class 
of currents is 
\be
\C \ = \ \tr \left [ (K^{\vphantom{\hc}} K^\hc)^s \right ] \ , 
\ee
and a second class is 
\be
\C = \tr \left [ \, (K^{\vphantom{\hc}} K^\hc)^{s-1} 
(K^{\vphantom{\hc}} L^\hc + L^{\vphantom{\hc}} K^\hc) \, 
\right ] 
\ee
They are non-zero only when $s$ is odd (by flipping indices on all
matrices, then re-arranging). 
The corresponding tensors are $\Omega^{(e)}$ 
and $\Lambda^{(d)}$, respectively, where $d$ and $e$ are both 
trace-powers (on $\g$ and $\h$). These currents are 
primitive for $s \leq n$, provided $n > 3$. (The case $n{=}3$
differs slightly from the general pattern, because of some special 
behaviour of the low-dimensional representations involved;
but this has already been treated 
above as one of the complex Grassmannians,
since $SO(6)/U(3) = SU(4)/S(U(3){\times}U(1)$.)

There is also a Pfaffian amongst the primitive $d$ invariants 
on $\frak{so}(2n)$, which gives rise to a $\Lambda$ tensor.
However, this tensor, and the corresponding current, are not
independent of those already written above. This is because
the $\varepsilon$ tensor of $\frak{so}(2n)$ which appears 
in the Pfaffian, is a product 
$\varepsilon_{i_1 \ldots i_n} \varepsilon^{j_1 \ldots j_n}$
in terms of $H = U(n)$ invariants. But such a product 
can be re-written in terms of $\delta_i{}^j$ tensors,
and hence in terms of traces.

The family $Sp(n)/U(n)$ is very similar.
We can again identify $A+iB\in \frak{u}(n)$ with the matrix 
$\begin{pmatrix} A & B \\ -B & A
\end{pmatrix}$ and elements of $\m$ take the form
${\textstyle{ i \begin{pmatrix} C & D  \\ D & - C 
\end{pmatrix} } }$ where $C$ and $D$ are now $n{\times}n$ real and
symmetric and can be combined in a complex symmetric 
matrix $C+iD$.
The invariants are the same as 
the trace-type invariants in the previous case, and the 
reasoning is very similar. 

\section{Overview of results obtained and those to follow}

\subsection{Summary of primitive currents}
The work of sections 3 and 4 determines, by
exhaustion, the primitive $\Lambda$ and $\Omega$ tensors for 
each classical symmetric space and we now 
summarize the results.
\[
\begin{array}{|c|c|}
\hline
& \\
G/H   &  \phantom{X} s: \ \Lambda^{(d)}~{\rm primitive},~d~{\rm of~degree}~
s{+}1\phantom{X}  
\\[0.5mm]
\hline
&  \\
\phantom{X}
SO(p{+}q)/SO(p){\times}SO(q) \ \ \ p\leq q \phantom{X}  
&  \! 1, \, 3, \, \ldots , \,
\left \{ 
\begin{matrix}
p{\phantom{-1}} \ \ \ \ p~{\rm odd} \\
p{-}1 \ \ \ \ p~{\rm even} \\
\end{matrix} \right .  \\  
& \quad \quad 
{\rm and}~{\frac{1}{2}}(p{+}q){-}1, \ \ p, \ q \ 
{\rm even} \phantom{XX} \hfil \\
& \\
\phantom{X}
SU(p{+}q)/S(U(p){\times}U(q)) \ \ \ p\leq q \phantom{X}  
&  1, \, 2, \, \ldots , \, p  \phantom{XXXx}  
\hfil \\
& \\
\phantom{XX}
Sp(p{+}q)/Sp(p){\times}Sp(q) \ \ \ \ \ p\leq q\phantom{XX}  
&  1, \, 3, \, \ldots , \, 2p{-}1 
\phantom{Xxi}  \hfil \\
& \\
SU(n)/SO(n) \ \ \ \ \ n \geq 2 \phantom{XXX} 
&  1, \, 3, \, \ldots , \,
\left \{ 
\begin{matrix}
n{-}2{\phantom{1}} \ \ \, n~{\rm odd} \\
n{-}1{\phantom{2}} \ \ n~{\rm even} \\
\end{matrix} \right .  \\
& \\
SU(2n)/Sp(n) \ \ \ \ \ n \geq 3 \phantom{XXX} 
&  1, \, 3, \, \ldots , \, 2n{-}1
\phantom{xxjXXX}\\
& \\
SO(2n)/U(n) \ \ \ \ \ n \geq 4 \phantom{XXX}
&  1, \, 3, \, \ldots , \,  
\left \{ 
\begin{matrix}
n{\phantom{-1}} \, \ \ \ \, n~{\rm odd} \\
n{-}1 \ \ \ n~{\rm even} \\
\end{matrix} \right . \\ 
& \\
Sp(n)/U(n) \ \ \ \ \ \ n \geq 2 \phantom{XXX}
&  1, \, 3, \, \ldots , \, 
\left \{ 
\begin{matrix}
n{\phantom{-1}} \ \ \ \, n~{\rm odd} \\
n{-}1 \ \ \ n~{\rm even} \\
\end{matrix} \right . 
\\[0.2in]
\hline
\end{array}
\]
The first table provides a list of primitive 
$\Lambda$ tensors and the integer $s$ is the spin of 
the top component, bosonic conserved charge. The first member
of each list has $s=1$, corresponding to the super-energy-momentum
tensor. The second table specifies the degrees
of the primitive cohomology generators, or $\Omega$ tensors
(these results could also be found using techniques such as those
in \cite{Milnor}-\cite{Spivak:vol5}).

\[
\begin{array}{|c|c|}
\hline
& \\
G/H   &  
\phantom{XX}{\rm degrees~of}~\Omega:~{\rm
cohomology~generators}\phantom{XX}
\\[0.5mm]
\hline
      &  \\
\phantom{X}
SO(p{+}q)/SO(p){\times}SO(q)  \ \ \ p\leq q \phantom{X}  
&  4, \, 8, \, \ldots , \,
\left \{ 
\begin{matrix}
2p{-}2 \qquad \qquad \, \ \ \ p~{\rm odd} \\
2p{-}4, \ \ {\rm and} \ \ p, \ \ p~{\rm even} \\
\end{matrix} \right .  \\  
& \phantom{XXXXXXXXXXX}  
{\rm and} \ \ q, \ \ q~{\rm even} 
\phantom{XX} \\
& \\
\phantom{X}
SU(p{+}q)/S(U(p){\times}U(q)) \ \ \ p\leq q \phantom{X}  
&  2, \, 4, \, \ldots , \, 2p  
\phantom{XXXXXXXXXXX}  
\\
& \\
\phantom{XX}
Sp(p{+}q)/Sp(p){\times}Sp(q)  \ \ \ \ \ p\leq q\phantom{XX}  
&  4, \, 8, \, \ldots , \, 4p
\phantom{XXXXXXXXXXX}  \\
& \\
SU(n)/SO(n) \ \ \ \ \ n \geq 2 \phantom{XXX}                   
&  \phantom{x}5, \, 9, \, \ldots , \,
\left \{ 
\begin{matrix}
2n{-}1{\phantom{1}} \, \ \ \ \ \ \ \ \ \ \ \ \ \ \  n~{\rm odd} \\
2n{-}3, \ \ {\rm and} \ \ n, \ \ n~{\rm even} \\
\end{matrix} \right .  \\
& \\
SU(2n)/Sp(n) \ \ \ \ \ n \geq 3 \phantom{XXX}                  
&  5, \, 9, \, \ldots , \, 4n{-}3 
\phantom{XXXXXXXXX} \\
& \\
SO(2n)/U(n) \ \ \ \ \ n \geq 3 \phantom{XXX}                  
&  2, \, 6, \, \ldots , \, 
\left \{ 
\begin{matrix}
2n{\phantom{-2}} \ \ \ \ \ \, n~{\rm odd} \\
2n{-}2 \ \ \ \ \ n~{\rm even} \\
\end{matrix} \right . \phantom{XXX} \\ 
& \\
Sp(n)/U(n) \ \ \ \ \ \ n \geq 2 \phantom{XXX}                   
&  2, \, 6, \, \ldots , \,  
\left \{ 
\begin{matrix}
2n{\phantom{-2}} \ \ \ \ \ \, n~{\rm odd} \\
2n{-}2 \ \ \ \ \ n~{\rm even} \\
\end{matrix} \right .  \phantom{XXX} \\[0.2in]
\hline
\end{array}
\]
Note that some of the families above resemble Lie groups in that
all (or almost all) their cohomology generators 
are of odd degree. The symmetric spaces with $\rank(\g)= \rank(\h)$, 
on the other hand, have all their cohomology generators of even degree.

\subsection{Poisson brackets}

Having arrived at a coherent description of the primitive classical 
conserved currents in each supersymmetric $G/H$ model, it is natural 
to ask about their Poisson bracket algebra.
Even for the case of groups, however, calculations of the full
classical current algebra are lengthy and involved \cite{SPCM}. 
We will therefore confine our attention to a 
particularly interesting aspect of this current algebra:
the Poisson brackets of the top component
charges arising from all the primitive currents. One reason these 
are of special significance is that it was shown in \cite{SPCM} that 
the primitive currents for the SPCMs could be chosen so that their 
bosonic, top component charges all Poisson-commute, and it is
natural to ask whether this generalizes to symmetric spaces.
A second reason is that it is the top component charges, and these
alone, that are most likely to survive quantization.
Although our results in this paper are purely
classical, possible future implications for the quantized theories 
are also a strong motivation for our work here. Let us recall why it is the
top component charges which are significant for the quantum theory. 

The superholomorphic form of the classical conservation laws (\ref{shol})
is intimately associated with the superconformal symmetry of the 
classical theory---which is why, of course, the conservation of the classical 
super-energy-momentum tensor itself takes this form,
$D_- ( \K^\am_+ {\cal D}_+ \K^\am_+) = 0$.
It is reasonable to anticipate that (at least some of) 
the local conservation laws we have found 
survive in the quantum theory \cite{GW,Clark,EKMY}. But since superconformal
invariance is broken quantum-mechanically, we should expect
that any conservation equations which survive will appear in 
the more general, modified form:
\be D_- \C = 
D_+ \widetilde{\C} \ ,
\label{nonhol}\ee
where $\widetilde{\C}$ is some new superfield, and $\C$ too may receive
corrections. 
It is easy to show, simply by introducing component 
expansions for $\C$ and $\widetilde{\C}$, that the equation (\ref{nonhol})
contains a generalization of the top component conservation equation
in (\ref{shol}), with charge $S$, but that there is, \emph{no}
generalization of the bottom component conservation law, in general
(for more details, see \cite{SPCM,EvansMadsen}). This is consistent
with the fact that the original top component charge $S$
and any quantum generalization of it are both invariant under supersymmetry.

We will denote by $F^{(e)}$ or $B^{(d)}$ the top component charges 
arising from a current $\C_\Omega$ based on 
forms $\Omega^{(e)}$ or $\Omega^{(d)}$ respectively; these charges are 
indeed fermionic and bosonic, as the notation
suggests. We will denote by $P^{(d)}$ the top
component charge arising from a current $\C_\Lambda$ based on
$\Lambda^{(d)}$; these are
always bosonic and the first member of the sequence is
the momentum, arising from 
the choice $\Lambda_{\am \bm} = \delta_{\am \bm}$.
Our aim in the remainder of this paper is to 
compute the Poisson bracket algebra of these quantities
defined using particular primitive tensors $d$ and $e$.
We will carry out the calculations in section 7, after 
analyzing the canonical structure of the sigma models in section 6,
but it is useful to state the results in advance as a guide to
the calculations which follow.

We will prove that for the special family of $d$ tensors obeying
(\ref{special}), all bosonic top component charges Poisson-commute:
\bea 
\{ B^{(d)} , B^{(\tilde d)} \} 
& = & 0 \label{BB} \\
\{ B^{(d)} , P^{(\tilde d)} \} 
& = & 0 \label{BP} \\
\{ P^{(d)} , P^{(\tilde d)} \} 
& = & 0 \label{PP}
\eea
We will also find that (graded) Poisson brackets 
involving the fermionic charges 
close in the following pattern:  
\bea 
\left\{ F^{(e)} , B^{(d)} \right\} & = & 0 \label{FB} \\
\left\{ F^{(e)} , P^{(d)} \right\} & = & \hat F  \label{FP} \\
\left\{ F^{(e)} , F^{(\tilde e)} \right\} & = & \hat P \label{FF}
\eea
Here $\hat F$ denotes a top component fermionic charge arising from a
current consisting of terms of the form 
$\C_{\Omega'} \del_+ \C_{\Omega''}$, while 
$\hat P$ denotes a bosonic charge arising from a current
$\C_\Lambda$ but with $\Lambda$ a compound tensor, in general.

It is certainly satisfactory that we can find
classically-commuting sets of bosonic charges in all 
models, generalizing the results of \cite{SPCM}.
The fermionic top component charges are a novel feature of the
symmetric space sigma models, however, with no counterparts in
the SPCMs.
It is worth emphasizing that we should \emph{not} expect to find 
sets of these fermionic charges with \emph{vanishing} Poisson brackets.
This is because the Poisson brackets are graded, so  
$\{ F , F \} = 0$ becomes the operator equation $F^2 = 0$ in the 
quantum theory, but if $F$ is hermitian this implies $F=0$,
and the charge is trivial (for a positive-definite Hilbert space). 
The algebra (\ref{FF}) can be 
regarded as some higher-spin generalization of supersymmetry.
We should also mention that a highly non-trivial check 
of our results on primitive currents is that the 
non-zero expressions found on the right hand sides of 
(\ref{FP}) and (\ref{FF}) have integrands which are indeed 
differential polynomials in the primitive currents we have identified.

\subsection{Hermitian symmetric spaces and ${\cal N} = 2$
supersymmetry}

It is a famous result that a (two-dimensional) supersymmetric sigma model
admits an additional supersymmetry iff its target space is a
K\"ahler manifold \cite{Alvarez}. The symmetric spaces $G/H$ which are 
K\"ahler, otherwise known as \emph{hermitian} symmetric spaces, 
are those for which $H$ is the product of $U(1)$
and some semi-simple factor \cite{Helgason}. We see from the second table 
in section 5.1 that in precisely these cases there is a cohomology generator 
of degree $2$, 
which is just the K\"ahler form.
The corresponding superfield current has spin-1, and so is not really 
`higher-spin' at all, but rather a conventional, Noether current. Its top 
component is fermionic and is
the spin-3/2 supercurrent for the second supersymmetry,
while its bottom, bosonic, component is the current for a chiral 
${\cal R}$-symmetry which rotates the supercharges in the ${\cal
N} = 2$ algebra into one another.

To elaborate on this, consider the component expansions of the 
K\"ahler holomorphic spin-1 current and its anti-holomorphic counterpart:
\be
r_+ \, + \, \theta^+ s_{+}  \qquad {\rm and} \qquad 
r_- \, + \, \theta^- s_{-}  
\ee
The conserved charges $R_{(\pm)}$ arising from $r_\pm$ are 
Lorentz scalars; they rotate 
the original supersymmetry charges, $Q_\pm$, into 
the additional supercharges, $S_\pm$, 
arising from the spin-3/2 currents $s_{\pm}$,
while leaving $Q_\mp$ unchanged. It is for this reason that we  
refer to these transformations as \emph{chiral} $R$-symmetries. 
These symmetries many be more familiar 
\cite{DAdda,Witten:77}
in the combinations 
$R = R_{(+)} + R_{(-)}$ and $\tilde R = R_{(+)} - R_{(-)}$
which correspond to the current conservation equations 
taken in the forms
\be
\del_- r_+ \ \pm \ \del_+ r_- = 0 \ .
\label{Rcurr}\ee
These currents are duals of one another (as vectors or
one-forms in two-dimensional Minkowski space),
the first being a vector, while the second is a pseudo-vector 
under Lorentz transformations together with reflections.

Note that the general form of the charge algebra given 
in (\ref{FF}) gives exactly what we expect when applied to the 
K\"ahler currents: the Poisson brackets of the new supercharges 
$S_\pm$ ($F$ type) with themselves yield energy-momentum
($P$ type).
It is also worth emphasizing that well-known properties
of certain ${\cal N} = 2$ models \cite{DAdda,Witten:77} confirm
our expectation that bottom component symmetries need not survive
quantization. In these models, the ${\cal N}=2$ supersymmetries
and the charge $R$, corresponding to the first combination in
(\ref{Rcurr}), all persist at the quantum level, but there
there is no conserved charge $\tilde R$.
The quantum violation of this symmetry is just the usual chiral 
anomaly, in two-dimensions.\footnote{There is actually a discrete remnant 
of the $\tilde R$ symmetry in the quantum theory and this is 
\emph{spontaneously} broken, which is important in understanding the
spectrum \cite{DAdda,Witten:77}.}

\section{Component fields and canonical structure}

\subsection{Component lagrangian}

To calculate Poisson brackets one needs the ordinary $x$-space form of the
superspace Lagrangian (\ref{G/Hlag}). Let us expand the 
basic $G$-valued superfield in term of real component fields as follows
\be 
\G ( x, \theta ) = g ( x ) \exp \left( i \theta^+ \psi_+ ( x ) + i
\theta^- \psi_- ( x ) + i \theta^+ \theta^- \mu( x ) \right) \ .
\label{Gcomp}
\ee
where $g (x) \in G$ and $\psi_\pm (x), \mu (x) \in \g$. 
The superfield currents defined in (\ref{Jdef}) are then
\bea
\J_+ & = & i \psi_+ 
\ - \ i \theta^+ \! \left( j_+ + i \psi^2_+\right) 
\ + \ i \theta^- \! \left( \, \mu - {\textstyle{\frac{i}{2}}} \left[
                           \psi_+, \psi_- \right] \, \right)\nn\\
&  & {} + \ \theta^+ \theta^- \! \left( \, 
\del_+ \psi_- + \left[ j_+, \psi_-\right] 
     + {\textstyle{\frac{i}{2}}} \left[ \psi^2_+, \psi_- \right] 
     + [\mu, \psi_+] \, \right) \ ,
\\ \label{compexp1}
\J_- & = & i \psi_- \ - \ i \theta^- \! \left( j_- + i\psi^2_-\right) 
\ + \ i \theta^+ \! \left( \, - \mu - {\textstyle{\frac{i}{2}}} \left[
                           \psi_-, \psi_+ \right] \, \right)\nn\\
         &  & {} + \ \theta^- \theta^+ \! \left( \, \del_- \psi_+ + 
\left[ j_-, \psi_+\right] 
                + {\textstyle{\frac{i}{2}}} 
\left[ \psi^2_-, \psi_+ \right] - [ \mu, \psi_-] \, \right) \ ,
\label{compexp2}
\eea
where we have introduced bosonic currents
\be
j_\pm = g^{-1} \del_\pm g \quad \in \ \ \g \ .
\label{jdef} \ee
The resulting component lagrangian for the $G/H$ sigma model is 
\bea L&=& {\textstyle{\frac{1}{2}}} \, 
( \, j_+ + i \psi^2_+\, )^\am(\, j_- + i \psi^2_- \,)^\am 
\ - \ {\textstyle{\frac{1}{2}}} \, 
(\, \mu - {\textstyle \frac{i}{2} } [\psi_+,\psi_-] \, )^\am
(\, -\mu - {\textstyle \frac{i}{2} } [\psi_-,\psi_+] \, )^\am \nn \\
 &&\quad\quad\quad
  + \ {\textstyle{\frac{i}{2}}} \,
\psi_+^\am ( \, \del_-\psi_+ +[j_-,\psi_+] + {\textstyle{\frac{i}{2}}}  
[\psi^2_-,\psi_+]-[\mu,\psi_-] \, )^\am \nn \\
 &&\quad\quad\quad
  + \ {\textstyle{\frac{i}{2}}} \, 
\psi_-^\am (\, \del_+\psi_- +[j_+,\psi_-] + {\textstyle{\frac{i}{2}}} 
[\psi^2_+,\psi_-]+[\mu,\psi_+] \, )^\am \ . \eea

This lagrangian, although written in terms of component fields, still
possesses the full superspace $H$-gauge symmetry, 
under which $\G \mapsto \G \HH$ and 
$\J_\pm = \K_\pm + \A_\pm$ transforms as in (\ref{Agauge}) and
(\ref{Kgauge}). It is best to partially fix this so as to leave 
only conventional, $x$-space, gauge-transformations. 
To see how this can be done, consider a component expansion
\be 
\HH (x, \theta) = h(x) \exp \! 
\left( \, i\theta^+ \eta_+(x) +i\theta^- \eta_-(x) 
+i\theta^+ \theta^- \nu(x) \, \right) \ee
with $h(x) \in H$ and $\eta_\pm (x), \nu (x) \in \h$.
It follows that under $\G \mapsto \G \HH$ we have 
\be
\psi_\pm^\am \ \mapsto \ ( \, h ^{-1} \psi_\pm h \, )^\am \ , 
\qquad \psi_\pm^\ah \ \mapsto \ ( \, h^{-1}\psi_\pm h + \eta_\pm \, )^\ah \ee
and so one can always impose $\psi_\pm^\ah=0$ by a unique choice of
$\eta_\pm^\ah$. Similarly, the gauge freedom inherent in $\nu^\ah$ can
be used up by setting $\mu^\ah=0$, and in fact $\mu^\am = 0$ then
follows from the lagrangian as an equation of motion.

This achieves the aim of reducing the gauge redundancy to 
$x$-space gauge transformations (the restrictions we have imposed 
constitute the \emph{Wess-Zumino gauge} for this superspace 
gauge theory, which can also be expressed by the condition 
$\theta^+ \A_+ + \theta^- \A_- = 0$).
Having done this, we have 
component expansions of the form 
\bea 
\K_\pm^\am & = & i\psi_\pm^\am \ - \ i\theta^\pm k_\pm^\am \ + \
O(\theta^\mp) 
\label{Kcomp}\\
\A_\pm^\ah & = &  - \ i\theta^\pm A_\pm^\ah \ + \ O(\theta^\mp)
\label{Acomp}
\eea
(higher components are functions of those given explicitly) where
\be
j _\pm  = k_\pm + A_\pm
\qquad {\rm with} \qquad 
k_\pm \in \m \ , \quad A_\pm \in \h \ . 
\label{jkacomp}\ee
Finally, then, the component formulation of the theory involves 
fields $g(x^\mu)$, and $\psi_\pm^\alpha$, with lagrangian
\bea L&=& {\textstyle \frac{1}{2} } \,  
( \, k_0^\am k_0^\am - k_1^\am k_1^\am \, )
\ + \ {\textstyle \frac{i}{2} } \, \psi_+^\am \del_- \psi_+^\am 
\ + \ {\textstyle \frac{i}{2} } \, \psi_-^\am \del_+ \psi_-^\am \nn \\ 
&&{}- \ i A_0^\ah ( h_+^2 + h_-^2 )^\ah
\ + \ iA_1^\ah ( h_+^2 - h_-^2 )^\ah 
\ + \ h^\ah_+ h^\ah_- \ , 
\label{complag}\eea 
where the various combinations of fields which appear 
are defined by (\ref{jdef}) and (\ref{jkacomp}) together with 
\be h_\pm^\ah \ = \ (\psi_\pm^2)^\ah \ = \ {\textstyle \frac{1}{2}} \,
f^{\ah \bm \gm} \psi_\pm^\bm \psi_\pm^\gm \ .
\label{hdef}\ee

\subsection{Poisson brackets}

We can now carry out a standard canonical analysis of the 
lagrangian in (\ref{complag}).  
All Poisson brackets are graded and hold at equal times, 
where appropriate.

The classical brackets for the real (Majorana) fermions are simply 
\be
  \left\{ \, \psi_\pm^\am ( x ), \, \psi_\pm^\bm ( y ) \, \right\} 
= - i \delta^{\am \bm} \dxy\label{psipbs} \ , 
\ee
with others vanishing. 
It is useful to note that these imply
\bea
\{ \, h_\pm^\ah(x), \, \psi_\pm^\bm(y) \, \} & = & 
if^{\ah \bm \gm} \psi_\pm^\gm(x) \dxy \\
\{ \, h_\pm^\ah(x), \, h_\pm^\bh(y) \, \} & = & 
if^{\ah \bh \gh} h_\pm^{\gh}(x) \dxy \ 
\label{halg}\eea
(with $h^\ah$ defined in (\ref{hdef})).

To calculate the brackets of the bosonic currents $j=k+A$ we introduce 
a set of coordinates $\{ \phi^i \}$ on the group manifold $G$ and
regard the field as depending on $x$ through these: 
$g(\phi^i (x))$. Vielbeins for the group will be denoted
\be
E^a_i (\phi) = ( g^{-1} \del_i g )^a \qquad \Rightarrow \qquad 
k_\pm^\am = E^\am_i \del_\pm \phi^i 
\ , \quad A_\pm^\ah = E^\ah_i \del_\pm \phi^i \ .
\ee
From (\ref{complag}), the momentum conjugate to $\phi^i$ is 
\be
\pi_i=\frac{\del L}{\del (\del_0 \phi^i)}
=E^{\am}_iE^{\am}_j\del_0\phi^j - iE^{\ah}_i\left(h_+ + h_-\right)^\ah
\ . \ee
The spatial components of the current $j$ can be expressed
\be 
j^a_1 = E^a_i(\phi) \del_1 \phi^i
\qquad \Rightarrow \qquad
k^\am_1 = E^\am_i(\phi) \del_1 \phi^i \ , \qquad 
A^\ah_1 = E^\ah_i(\phi) \del_1 \phi^i
\ee
and it is convenient to define a new current, related 
to momentum:
\be
J^a = E^{ai} (\phi) \pi_i \qquad \Rightarrow \qquad 
J^\am = k_0^\am \ , 
\qquad J^\ah = -i(h_+ + h_-)^\ah \ .
\label{Idef}\ee
The Poisson brackets of all these quantities can now be calculated
from 
\be \left\{ \, \phi^i(x), \, \pi_j(y) \, \right\} \ 
= \ \delta^i{}_j \, \dxy \ee
(similar calculations are described in appendices to \cite{PCM,SPCM}). 
The results are:
\bea 
\left\{ \, J^a ( x ), \, J^b ( y ) \, \right\} & = & 
- f^{a b c} J^c ( x ) \dxy\nn\\
\left\{ \, J^a ( x ), \, j_1^b ( y ) \, \right\} & = & 
- f^{a b c} j_1^c ( x )\dxy \ + \ \delta^{a b} \dpxy \nn\\
\left\{ \, j_1^a ( x ), \, j_1^b ( y ) \, \right\} & = & \ \ 0 \ .
\label{currpbs}\eea
and the brackets of $J^a$ and $j_1^a$ with the fermions $\psi_\pm^\am$
all vanish.

However, one of the formulas in (\ref{Idef}) constitutes a 
constraint on the canonical variables 
\be 
\I^\ah = J^\ah + i\left(h_+ + h_-\right)^\ah \approx 0
\label{constraintsusy}\ee
The Hamiltonian density for the system is, therefore, 
only weakly determined 
\be 
H \ \approx \ {\textstyle \frac{1}{2}} k_0^\am k_0^\am 
+ {\textstyle \frac{1}{2}} k_1^\am k_1^{\am} 
- i A_1^\ah ( h_+ - h_- )^\ah
+ {\textstyle \frac{1}{2}} i \psi_+^\am \del_1 \psi_+^\am 
- {\textstyle \frac{1}{2}} i \psi_-^\am \del_1 \psi_-^\am 
- h_+^\ah h_-^\ah \ .\ee
From the brackets (\ref{currpbs}) and (\ref{psipbs}) it is easy to
show that 
\be
  \left\{ \, H, \, \I^{\ah} \, \right\} \approx 0 \ , 
\ee
so that there are no secondary constraints, and also
\be
\{ \, \I^{\ah} (x) , \, \I^{\bh} (y) \, \} = -f^{\ah \bh \gh} 
\I^{\gh} (x) \delta(x{-}y) \ , 
\ee
from (\ref{currpbs}) and (\ref{halg}), 
so that the constraints are first class. 
The canonical formalism has thus been consistently completed.

The general nature of these results is expected: the
lagrangian (\ref{complag}) has an $x$-space $H$-gauge symmetry,
which is generated by the first-class constraints $\I^{\ah}$.
It is important to emphasize that we will \emph{not} fix this gauge
symmetry in any of the calculations which follow. To do so would
require the introduction of Dirac brackets, which would 
certainly differ from the Poisson brackets above, in general.
However, we will be interested, ultimately in the 
Poisson brackets of the gauge-invariant charges of type
$B$, $F$ and $P$, and Poisson brackets and Dirac brackets 
coincide for any gauge-invariant quantities.

\subsection{Component forms of the charges}
From the component expansion of the current $\K_+$ given in
(\ref{Kcomp}), it is easy to see that the top component charges
associated with $\Omega$ tensors are (up to irrelevant, overall factors)
\bea 
B^{(d)}  \ & = & \  
\int \! dx \, \Omega^{(d)}_{\am_1 \dots \am_{2s} \am} \,
\psi_+^{\am_1} \dots \psi_+^{\am_{2s}} k_+^\am 
\label{Bcomp} \\
F^{(e)}  \ & = & \  
\int \! dx \, \Omega^{(e)}_{\bm_1 \dots \bm_{2s+2}} \,
\psi_+^{\bm_1} \dots \psi_+^{\bm_{2s+1}} k_+^{\bm_{2s+2}} 
\label{Fcomp}
\eea
and they have spins $s$ and $s+1/2$ respectively, 
where $s$ is an integer and $s{+}1$ is the degree of the symmetric
tensor $d$ or $e$.

The top component charges associated with $\Lambda$ tensors 
look rather more complicated. First note that 
\be \D_+ \K_+^\am  \ = \ -i k_+^\am \ + \ 
\theta^+(\, \del_+\psi_++[A_+,\psi_+] \, )^\am \ + \ O(\theta^-)
\ ,\ee 
and that to express this in terms of good canonical variables it is
necessary to eliminate $\del_0 \psi_+$ using the equation of motion
\be 
\del_- \psi_+ = - \, [ \, A_- + ih_-, \, \psi_+ \, ] 
\qquad \Rightarrow \qquad  
\del_+ \psi_+ = - \, [ \, A_- + ih_-, \, \psi_+ \, ] \ + \ 
2 \psi_+'^\am
\ee
(this is most easily derived from $\D_- \K_+ =0$, rather than using the
component lagrangian). It follows that 
\be  \label{DJ} 
\D_+ \K_+^\am   =  -i k_+^\am \ + \ \theta^+
\left(2\psi_+'+[2A_1-ih_-,\psi_+]\right)^\am+O(\theta^-) \ee
and hence
\be
P^{(d)} = 
-\int \! dx \, \Lambda_{ \am_1 \dots \am_{2s-1} ; \am}^{(d)} 
\left[ 
      (2s{-}1) k_+^\am k_+^{\am_{2s-1}} 
            - i ( 2 \psi'{}_+^\am - f^{\am \bh \gm} \X^\bh \psi_+^\gm )
\psi_+^{\am_{2s-1}}
\right] 
\psi_+^{\am_1} \ldots \psi_+^{\am_{2s-2}} \label{Pcomp} \ee
is the charge of spin $s$ resulting from a $d$ tensor of degree $s+1$,
where 
\be \X^\ah = i(h_+ +h_-)^\ah -2A_1^\ah \ .
\label{Xdef}\ee
We have used invariance of $\Lambda$ to include 
the $h_+$ term in this definition.
This proves convenient because 
$\X$ is then the quantity which appears in the bracket of $k_+$ with itself:
\be\left\{ k_+^\am(x),k_+^\bm(y) \right\} \ = \ 
f^{\am \bm \gh} \X^\gh(x)\dxy
                  +2 \delta ^ {\am \bm} \dpxy \ , 
\label{kpbs}\ee
which follows from (\ref{currpbs}). 

We now have all the information we 
need and we proceed to compute Poisson brackets and 
establish the results announced in 
section 5.

\section{Calculations of Poisson Brackets of charges}

\subsection{Brackets amongst $B$ and $F$ type charges}

Consider the bracket of two charges of type (\ref{Bcomp}) or
(\ref{Fcomp}) 
based on tensors $\Omega_{\am_1 \am_2 \dots \am_{p} }$ 
and $\widetilde\Omega_{\bm_1 \bm_2 \dots \bm_{q}}$ where, for the 
moment, $p$ and $q$ can be even or odd integers, so that the charges 
can be type $B$ or $F$.
From (\ref{kpbs}) and the fact that $k_+^\am$ and $\psi_+^\bm$
Poisson-commute we find the result:
\bea
&&(q-1) \int \! dx \, \Omega_{\am_1 \dots \am_{p-2} \am_{p-1} \gm} \,
\widetilde\Omega_{\bm_1 \bm_2 \dots \bm_{q-1} \gm} 
\psi_+^{\am_1} \dots \psi_+^{\am_{p-2}} 
\psi_+^{\bm_2} \dots \psi_+^{\bm_{q-1}} \nn\\&&\qquad\qquad
\times \left[ -i(p{-}1)k_+^{\am_{p-1}} k_+^{\bm_1} + \psi_+^{\am_{p-1}} 
                  ( 2 \psi'{}_+^{\bm_1} - f^{\bm_1 \gh \bm} 
\X^\gh \psi_+^\bm ) \right] \ , \label{laminta} 
\eea
which is obtained after integration by parts and using the
invariance of $\widetilde\Omega$. Taking into account various factors
arising from antisymmetrization of tensor indices, this integrand 
is proportional to
\bea&& 
\Omega_{[\am_1 \dots \am_{p-1} }{}^\gm  \, 
\widetilde\Omega_{\bm_1 \dots \bm_{q-2} ] \bm \gm}\;
 \psi_+^{\am_1} \dots \psi_+^{\am_{p-1}} \,
\psi_+^{\bm_1} \dots \psi_+^{\bm_{q-3}}\nn\\
&&\quad \quad\times \left[ (p{+}q{-}3) 
k_+^\bm  k_+^{\bm_{q-2}} - i \psi_+^{\bm_{q-2}}
( 2\psi'{}_+^\bm - f^{\bm\gh\delta}\X^\gh\psi_+^\delta ) \right] \ .  
\label{lamint}\eea
Because we have taken the bracket of two conserved charges,
the result must be another conserved quantity. We should therefore be
able to express the integrand above in terms of our known, primitive 
conserved currents. It clearly has the general form required for a
charge of type $P$ given by (\ref{Pcomp}), if the $\Lambda$ tensor 
is taken as 
\be
\Omega_{[\am_1 \dots \am_{p-1} }{}^\gm \, 
\widetilde\Omega_{\bm_1 \dots \bm_{q-2} ] \bm \gm} \ .
\label{OO}\ee
But we must check that such an expression really can arise from
our construction of $\Lambda$ tensors given in sections 3 and
4 (if not, we would have found a new conserved quantity, not
expressible in terms of the currents we claim are primitive).

In fact, if either one of the original tensors, say $\widetilde\Omega$, 
is of odd degree, then (\ref{lamint}) actually \emph{vanishes}.
The crucial point is that for  
\be \widetilde\Omega_{\am_1 \dots \am_{2s+1}} = \frac{1}{2^{s}} 
f^{\roh_1}_{\am_1 [\am_2} \dots 
f^{\roh_s}_{\am_{2s-1} \am_{2s}]} d_{\am_{2s+1} \roh_1 \dots \roh_s}\ , 
\nn\ee
(referring to (\ref{Omegag}) and setting $q=2s+1$) 
the explicit antisymmetrization is only necessary over $2s{-}1$ of
the indices. When this is substituted into (\ref{lamint}), all the required
antisymmetrization is enforced by the presence of the fermions $\psi_+$. 
The whole integrand then vanishes by 
invariance of the tensor $\Omega$ (the term with $\psi_+'$
immediately, and the $k_+k_+$ term because invariance
of $\Omega$ produces an expression with both bosons $k_+$ 
contracted with the antisymmetric $\widetilde\Omega$). 
This establishes the relations $\{ B^{(d)}, B^{(\tilde d)} \} = 
\{ B^{(d)} , F^{(e)} \} = 0$.

More interesting is the result for the remaining kind of bracket, 
$\{ F^{(e)}, F^{(\tilde e)} \}$, which requires 
both $\Omega$ and $\widetilde\Omega$ to be of even degree.
We will show that when $e$ and $\tilde e$ are single trace
invariants or Pfaffians, the tensor (\ref{OO}) is indeed
a known $\Lambda$ tensor (possibly composite, or even zero in some 
cases). The generalization to products of traces and Pfaffians 
presents no difficulties of principle, the arguments would just be 
more cumbersome to write down.

Let us deal with the trace-type $e$ invariants first, which occur 
for the Grassmannians and for the families $SO(2n)/U(n)$ and $Sp(n)/U(n)$.
These symmetric spaces are defined by an automorphism on $\g$ of 
the form $X \mapsto N X N$ for some matrix $N$. Referring to the
block forms introduced in section 4, we have $N= 
\begin{pmatrix}1 & 0 \\ 0  & -1 \end{pmatrix}$ for the Grassmannians
and $N= \begin{pmatrix}0 & i \\ -i  & 0 \end{pmatrix}$ for the 
remaining two families. By definition, 
$N$ commutes with generators $t_{\ah}$ belonging to $\h$ but 
anticommutes with generators $t_\alpha$ belonging
to $\m$. 

Recall that the tensors $e$ are {\em not} the restrictions of
invariant tensors on $\g$. Nevertheless, 
those involving a single trace can be expressed 
\be 
e_{\ah_1 \ah_2 \ldots \ah_k} = {\rm Tr} \left( N t_{( \ah_1} t_{\ah_2} 
\ldots t_{\ah_k )} \right) 
\ee
(the additional factor of $N$ in the trace prevents 
this from being the restriction to $\h$ of a quantity like (\ref{gtr})).
The corresponding antisymmetric tensor can be written
\be \Omega_{\am_1 \dots \am_{2k-1} \am_{2k}} = 
\tr (N t_{[\am_1} \ldots
t_{\am_{2k-1}} t_{\am_{2k}]}) =
\tr (N t_{[\am_1} \ldots
t_{\am_{2k-1}]} t_{\am_{2k}})
\ee
and the fact that the last expression is automatically totally
antisymmetric will be important below.
In addition, whenever this $\Omega$ tensor is non-zero, we have 
\be 
N t_{[\am_1} t_{\am_2} \ldots t_{\am_{2k-1}]}
= - t_\gm \, \tr ( \, N t_{[\am_1} t_{\am_2} \ldots t_{\am_{2k-1}]}
t_{\gm} \, ) 
= - t_\gm \Omega_{\am_1 \am_2 \ldots \am_{2k-1} \gm}  \ .
\ee
This is a consequence of the fact that the matrix 
on the far left actually belongs to $\m$, or in some cases
the complexification of $\m$, and so can be expanded as a (real or 
complex) linear combination of generators. It is straightforward to 
check this claim e.g.~using the block forms given in section 4.

We now return to (\ref{OO}) for the case of two even-degree tensors 
based on single traces.
Setting $p = 2k$ and $q= 2\ell$, we have 
\bea
&&
\Omega_{\am_1 \dots \am_{2k-1} \gm}  \;
\widetilde\Omega_{\bm_1 \dots \bm_{2 \ell -2} \bm \gm} \nn \\
&=&
\tr (\, N t_{[\am_1} \dots t_{\am_{2k-1}]} t_\gm \, ) 
\, \tr (\, t_\gamma N t_{[\bm_1} \dots t_{\bm_{2 \ell -2}} t_{\bm]} \,) 
\nn\\ 
&=& 
- \tr (\, N t_{[\am_1} \dots t_{\am_{2k-1} ] } 
N t_{[ \bm_1} \dots t_{\bm_{2\ell-2}} t_{\bm ]}  
\, )  
\nn\\
&=& 
\tr (\, t_{[\am_1} \dots t_{\am_{2k-1}]} t_{[\bm_1} \dots t_{\bm_{2\ell-2}} 
t_{\bm ]} \, )  \nn
\eea
and on imposing the correct antisymmetrization,
\bea
&& \Omega_{[\am_1 \dots \am_{2k-1} }{}^\gm  \;
\widetilde\Omega_{\bm_1 \dots \bm_{2 \ell -2} ] \bm \gm} \nn \\
& = &
\tr (\, t_{[\am_1} \ldots t_{\am_{2k-1}} t_{\bm_1} \dots t_{\bm_{2\ell-2}]} 
t_{\bm } \, )  \nn \\
& \propto &
f^{\gh_1}{}_{[\am_1 \am_2} \ldots f^{\gh_{k}}{}_{\am_{2k-1} \bm_1} 
\ldots f^{\gh_{\ell+k-1}}{}_{\bm_{2\ell -4} \bm_{2\ell -3}}
       \tr ( t_{\bm_{2\ell -2]}} t_\bm t_{\gh_1} 
\ldots t_{\gh_{\ell+k-1}} ) \nn\\ 
& \propto &
f^{\gh_1}{}_{[\am_1 \am_2} \ldots f^{\gh_{k}}{}_{\am_{2k-1} \bm_1} 
\ldots f^{\gh_{\ell+k-1}}{}_{\bm_{2\ell -4} \bm_{2\ell -3}}
       d_{\bm_{2\ell -2}] \bm \gh_1 \ldots \gh_{\ell +k-1}} \nn\\ 
  & \propto &
\Lambda^{(d)}_{\am_1 \am_2 \ldots \am_{2k-1} 
\bm_1 \bm_2 \ldots \bm_{2\ell-2} ; \beta}
\eea
where $d$ is the symmetric $G$-invariant on $\g$ associated 
with the invariant polynomial $\tr X^{k+\ell +1}$.
This is indeed one of our standard tensors, although neither $d$ 
nor $\Lambda$ need be primitive.
As mentioned above, this analysis can be extended to products of
traces.

The even-degree $\Omega$ tensors we have not yet considered
are constructed from Pfaffians. They occur only for the real
Grassmannians and for the spaces $SU(2n)/SO(2n)$. 
The special nature of the Pfaffian invariants  
means that it is convenient to adopt a notation similar to 
that used in (\ref{pf1})-(\ref{pf3}) and (\ref{pf4}).
We shall not give a detailed account of the calculations involving
these charges, but we will show how the final result is once again 
based on a $\Lambda$ tensor (possibly composite) of known form.

Considering first the Grassmannians $SO(p{+}q)/SO(p){\times}SO(q)$, 
we can regard $(\psi_+)_{i \ell}$ and $(k_+)_{i \ell}$ 
as $p{\times}q$ matrices 
transforming under $SO(p)$ from the left and $SO(q)$ from the right,
as in (\ref{haction}). Suppose $p$ is even, so we have a Pfaffian current
(\ref{pf1}) in addition to trace-type currents (\ref{grasstr}).
Setting $p=2n$, it is sufficient for our purposes to investigate the 
bracket of 
the \emph{bottom} component of the Pfaffian current, proportional to
\be 
\varepsilon_{i_1 j_1 \dots i_{n} j_{n}} (X)_{i_1 j_1} \ldots 
(X)_{i_n  j_{n}} \qquad {\rm where} \qquad X_{ij} = 
(\psi_+)_{i \ell} (\psi_+)_{j \ell} 
\label{botpfaff}\ee
with the top component of the trace current, proportional to 
\be
\tr (Y X^{r-1}) \qquad {\rm where} \qquad Y_{ij} = 
(\psi_+)_{i \ell} (k_+)_{j \ell}  
\label{toptrace}\ee
and to identify the result as the \emph{bottom} component
of a superfield current based on some $\Lambda$ tensor.
The relationship between the top component charges then follows
by supersymmetry (under which bottom component charges transform 
into top component charges, while top component charges are unchanged). 

A short calculation reveals that 
the Poisson bracket of the quantities
(\ref{botpfaff}) and (\ref{toptrace}) produces an expression
proportional to 
\be
\varepsilon_{i_1 j_1 i_2 j_2 \dots i_{n} j_{n}} 
(Y X^{r-1})_{i_1 j_1} X_{i_2 j_2} \ldots 
X_{i_{n}  j_{n} } 
\ee
The crucial point now is that 
\be
\varepsilon_{i_1 j_1 i_2 j_2
\ldots 
i_n  j_n} X_{k j_1}  
X_{i_2 j_2} \ldots 
X_{i_n j_n } \ =  \  2^{n-1}(n{-}1)! \, {\rm Pf} (X)
\, \delta_{i_1 k}
\ee
for {\em any} antisymmetric matrix $X$. Hence the expression above
factorizes into a Pfaffian and a trace and we recognize this as the
bottom component of a superfield current which is a product of 
terms (\ref{grasstr}) and (\ref{grasskl}). 

If both $p$ and $q$ are even, then there are Pfaffian 
currents (\ref{pf1}) and (\ref{pf2}) corresponding to both 
$SO(p)$ and $SO(q)$. Proceeding similarly, it is not 
difficult to see that the Poisson bracket of the 
top component $F$ charges gives a $P$ charge based on 
a $\Lambda$ tensor built from the Pfaffian of $SO(p{+}q)$.
Finally, we can consider the bracket of a Pfaffian $F$ charge with
itself, either in one of the real Grassmannians or in $SU(2n)/SO(2n)$
(in the latter case this is the unique $F$ charge).
Given the identity
\be \varepsilon_{i_1 i_2 \dots i_{2n-1} i_{2n}} 
\varepsilon^{\,j_1 j_2 \dots j_{2n-1} j_{2n}} = (2n)!\, 
\delta^{j_1}_{[i_1} \delta^{j_2}_{i_2} 
\dots \delta^{j_{2n-1}}_{i_{2n-1}} \delta^{j_{2n}}_{i_{2n}]},\ee
it is straightforward to verify that the result is a sum of 
terms, each a product of 
factors of type (\ref{grasstr}) together with a 
single factor of type (\ref{grasskl}).

This completes the argument that the brackets of $F$ charges 
close onto $P$ charges.

\subsection{Brackets amongst $P$ type charges}
We turn now to computing the bracket of two charges of type
$P^{(d)}$, based on tensors of type $\Lambda^{(d)}$.
We will show that, for special choices of the tensors $d$,
these charges actually Poisson-commute.
The calculations are rather involved, but we can follow the 
same strategy as in \cite{SPCM}.
Setting $s=m+1$ in (\ref{Pcomp}), it is useful to write 
\be
P^{(d)} = - \int dx \, d_{\am \bm \gh_1 \dots \gh_m}
\left[
(m+1) k_+^\am k_+^\bm 
- i ( 2\psi'{}_+^\am - f^{\am \bh \gm} \X^\bh \psi_+^\gm ) \psi^\bm_+
\right] h_+^{\gh_1} \ldots h_+^{\gh_m},
\ee
which follows on using the $G$-invariance of $d$. Hence 
\be P^{(d)} = -U - V - W ,\ee
where 
\bea
U &=& 2i\int \! dx \, d_{\am \bm \gh_1 \dots \gh_m} 
\psi_+^\am \psi'{}_+^\bm h_+^{\gh_1} \ldots h_+^{\gh_m},\\
V &=& (m{+}1) \! \int \! dx \, d_{\am \bm \gh_1 \dots \gh_m} 
k_+^\am k_+^\bm h_+^{\gh_1} \ldots h_+^{\gh_m},\\
W &=& i\int \!dx \, d_{\am \bm \gh_1 \dots \gh_m} 
f^{\am \deh \eps} \X^\deh \psi_+^\eps \psi_+^\bm  
h_+^{\gh_1} \ldots h_+^{\gh_m} \ ,\nn \\
&=& 2i \int \! dx \, d_{\deh \gh_1 \dots \gh_{m+1}} \X^\deh 
h_+^{\gh_1} \ldots h_+^{\gh_{m+1}} \ ,
\eea
(the last equality again relies on invariance of $d$).
We will compute the Poisson bracket of $P^{(d)}$ with a second charge,
$P^{(\tilde d)}$, expressed similarly as
\be 
P^{(\tilde d)} = - {\tilde U} - {\tilde V} - {\tilde W}
\ee
with $s = l + 1$.
This will involve consideration of various groups of terms arising from 
the brackets of $U$, $V$ and $W$ with $\tilde U$, $\tilde V$ and
$\tilde W$.

First, $\{W, \tilde W \}$ can be shown to vanish by repeated use of invariance 
of $d$ and $\tilde d$. Turning next to $\{ U, \tilde W \} + \{ W,
\tilde U \}$, we find that 
\be  \label{Uh}\{U,h_+^\gh(x)\}= -4 d_{\gh \ah_1 \dots \ah_{m+1} } 
\del_x ( 
h_+^{\ah_1} \ldots h_+^{\ah_{m+1}} ) \ , 
\ee
after extensive use of invariance of $d$, and hence
\be
\{U, \tilde W \}+\{W, \tilde U \} 
= -8i\int dx \, d_{\gh \ah \ah_1 \dots \ah_m} 
\tilde d_{\gh \bh \bh_1 \dots \bh_l}
h_+^{\ah_1} \ldots h_+^{\ah_m} h_+^{\bh_1} \ldots h_+^{\bh_{l}} 
\left( h'{}_+^\ah \X^\bh - h'{}_+^\bh \X^\ah \right) \label{UW}
\ee
This can be shown to vanish as follows. The antisymmetry in 
$\ah \leftrightarrow \bh$ imposed by the 
last factor allows us to symmetrize over the indices 
$(\ah, \ah_1, \dots, \ah_m, \bh_1, \dots, \bh_l)$ 
on the $d$ tensors. But $d_{\gamma \ah \ah_1 \ldots
\ah_{m}} = 0$, from (\ref{grade1}),
so that the repeated index $\gh$ in
(\ref{UW}) can be extended to a repeated index $c$, running over the 
whole Lie algebra $\g$. The tensor structure in (\ref{UW}) is therefore
\be
d_{c (\ah \ah_1 \dots \ah_m} 
\tilde d^{c}{}_{\bh_1 \dots \bh_l) \bh}
=
d_{c (\ah \ah_1 \dots \ah_m} 
\tilde d^{c}{}_{\bh_1 \dots \bh_l \bh)} \ ,
\ee
\emph{provided} the tensors $d$ and $\tilde d$ 
have the special property (\ref{special}).
For these special choices, then,
the symmetrization is actually over \emph{all} the indices 
$(\ah, \bh, \ah_1, \dots \ah_m,
\bh_1,\dots,\bh_l)$, but this implies
that the contraction with the rest of the integrand in (\ref{UW})
gives zero, because the last factor is antisymmetric in $(\ah, \bh)$.

Now consider the terms 
\bea
\left\{ V, \tilde V \right\} &=& 
(m{+}1)(l{+}1)\int dx \, 
d_{\am \bm \gh_1 \dots \gh_m} 
\tilde d_{\de \eps \roh_1 \dots \roh_l} 
\Big( 4 f_{\am \de \kh} \X^\kh k_+^\bm k_+^\eps h_+^{\gh_1} \ldots h_+^{\gh_m} 
                                        h_+^{\roh_1} \ldots h_+^{\roh_l} \nn\\
&& \quad\qquad\qquad\qquad + \, iml \, k_+^\am k_+^\bm k_+^\de k_+^\eps  
f_{\gh_1 \roh_1 \kh} h_+^\kh  h_+^{\gh_2} \ldots h_+^{\gh_m} 
         h_+^{\roh_2} \ldots h_+^{\roh_l} \Big) +\dots\\
\left\{V, \tilde W \right\} 
&=& 2i(m{+}1)\int dx \, d_{\am \bm \gh_1 \ldots \gh_m} 
\tilde d_{\deh \roh_1 \ldots \roh_{l+1}}
\Big(+4 f_{\am \deh \kappa} k_1^\kappa k_+^\bm h_+^{\gh_1} \ldots h_+^{\gh_m} 
                 h_+^{\roh_1} \ldots h_+^{\roh_{l+1}} \nn\\
&& \quad\qquad\qquad\qquad
+\, i m(l{+}1) \, f_{\gh_1 \roh_1 \kh} h_+^\kh k_+^\am k_+^\bm 
\X^\deh h_+^{\gh_2} \ldots h_+^{\gh_m} h_+^{\roh_2} \ldots 
h_+^{\roh_{l+1}}\Big) \\
\left\{W , \tilde V \right\} 
&=& -2i(l{+}1)\int dx \, d_{\deh \gh_1 \dots \gh_{m+1}} 
\tilde d_{\am \bm \roh_1 \ldots \roh_{l}} 
\Big(+4 f_{\am \deh \kappa} k_1^\kappa k_+^\bm h_+^{\gh_1} \ldots 
h_+^{\gh_{m+1}} 
h_+^{\roh_1} \ldots h_+^{\roh_{l}} \nn\\
&& \quad\qquad\qquad\qquad
+\, i l( m{+}1) \, f_{\roh_1 \gh_1 \kh}
h_+^\kh 
k_+^\am k_+^\bm 
\X^\deh h_+^{\gh_2} \ldots h_+^{\gh_{m+1}} 
h_+^{\roh_2} \ldots h_+^{\roh_{l}}\Big)
\eea
where we have neglected the non-ultra-local contributions (involving
$\delta'$) to $\{V, \tilde V \}$ 
for the moment. The $k_+k_+k_+k_+$ term may be shown to vanish by
repeated use of 
invariance of $d$ and $\tilde d$.
Now observe that, again by invariance of $d$ and $\tilde d$,
\be
0= 4f_{a_1 b_1 \kh} \X^\kh d_{a_1 \dots a_{m+2}} \tilde d_{b_1 \dots
  b_{l+2}} \eta_{a_2} \ldots \eta_{a_{m+2}} \eta_{b_2} \ldots
  \eta_{b_{l+2}} 
\Big|_{\lambda^2},
\ee
where
\be \eta = h_+ + \lambda k_+ ,\ee
and in fact when this expression is expanded out in terms of $h_+$'s 
and $k_+$'s, it produces precisely 
the $k_+k_+$ terms in $\{V, \tilde V\} + \{W, \tilde V\} + \{ V,
\tilde W\}$.
The remaining terms in $\{W, \tilde V\} + \{ V, \tilde W\}$ may be
written, up to a factor, as
\bea d_{a c_1 \dots c_{m+1}} \tilde d_{b d_1 \dots d_{l+1}} 
f_{a b \kappa} k_1^\kappa \eta^{c_1} \ldots \eta^{c_{m+1}} 
\eta^{d_1} \ldots \eta^{d_{l+1}} \Big|_{\lambda}
\eea
which also vanishes.

Thus, all the ultra-local terms in $\{V, \tilde V\} + \{W, \tilde V\} + 
\{ V, \tilde W\}$ vanish. The only 
non-ultra-local contribution comes from $\{V, \tilde V\}$ and is 
\bea
&& 8(m{+}1)(l{+}1)\int dx \, d_{\am \bm \gh_1 \dots \gh_m} 
\tilde d_{ \am \eps \roh_1 \dots \roh_l} 
                          k_+^\bm h_+^{\gh_1} \ldots h_+^{\gh_m} 
\del_x ( k_+^\eps h_+^{\roh_1} \ldots h_+^{\roh_l} ) \ . 
\eea 
Meanwhile, using the expression obtained previously for $\{U,h_+\}$, we have
\beaa
\left\{ U, \tilde V \right\}
 &=&  4l(l{+}1) \int dx \, d_{\deh \ah_1 \dots \ah_{m+1}} 
\tilde d_{\deh \am \bm \roh_1 \ldots \roh_{l-1}}
h_+^{\ah_1} \ldots h_+^{\ah_{m+1}}                         
\del_x (k_+^\am k_+^\bm  h_+^{\roh_1} \ldots h_+^{\roh_{l-1}} ) 
\eeaa
and
\beaa
\left\{ V, \tilde U \right\} 
&=& 4m(m{+}1) \int dx \, d_{\deh \am \bm \roh_1 \dots \roh_{m-1}} 
\tilde d_{\deh \ah_1 \dots \ah_{l+1}} 
k_+^\am k_+^\bm  h_+^{\roh_1} \ldots h_+^{\roh_{m-1}}
\del_x ( h_+^{\ah_1} \ldots h_+^{\ah_{l+1}} ) \ .
\eeaa
The sum of these last three terms is
\bea
&& 8 \int \! dx \, 
d_{a b_1 \dots b_{m+1}} \tilde d_{a c_1 \dots c_{l+1}} \eta^{b_1} 
\ldots \eta^{b_{m+1}} 
\del_x \left( \eta^{c_1} \ldots \eta^{c_{l+1}} \right) \Big|_{\lambda^2} \\
&\propto& \int \! dx \, d_{a (b_1 \dots b_{m+1}} \tilde d^a{}_{c_1 \dots
  c_l) c_{l+1}} 
                      \eta^{b_1} \ldots \eta^{b_{m+1}} 
\eta^{c_1} \ldots \eta^{c_l} \del_x \eta^{c_{l+1}} \Big|_{\lambda^2}.
\eea
Now we can invoke the property (\ref{special}) and extend the
symmetrization over all the
indices $(b_1,\dots,c_{l+1})$. The integrand is then a total
derivative, and so the whole expression 
vanishes.

There is only one remaining piece to consider in the bracket of
$P^{(d)}$ with $P^{(\tilde d)}$, 
namely $\{U, \tilde U\}$. This can be shown to be proportional to
\be d_{\am \bm \gh_1 \dots \gh_m}\tilde 
d_{\am \eps \roh_1 \dots \roh_l} h_+^{\gh_1} \ldots h_+^{\gh_m}
h_+^{\roh_1} \ldots h_+^{\roh_l} \psi'{}_+^\bm \psi'{}_+^\eps, 
\label{target}\ee
which again vanishes given the property (\ref{special}) for $d$ and $\tilde d$.
This completes the argument that
\be \{ P^{(d)}, P^{(\tilde d)} \} = 0 \ee
for the special set of primitive invariants $d$ defined in
(\ref{special}). 

\subsection{Brackets of $B$ or $F$ charges with 
$P$ charges}
We now consider the Poisson brackets between top component charges based on
$\Omega$ tensors (odd or even degree) and those based on $\Lambda$ tensors.

First, the bracket 
\be \left\{B^{(\tilde d)}, P^{(d)} \right\} 
= -\left\{ B^{(\tilde d)}, U + V + W \right\} \ .\ee
can be found by calculations similar to those in the last section.
Given (\ref{Uh}), we find
\be \left\{ U, B^{(\tilde d)} \right\} = -4 (m{+}1)(l{+}1) 
\! \int \! dx \, 
d_{\deh \gh_1 \ldots \gh_{m+1} } \tilde d_{\bm \bh_1 \ldots \bh_l \deh}  
\, k_+^\bm h_+^{\bh_1} \ldots h_+^{\bh_l} h_+^{\gh_1} 
\ldots h_+^{\gh_m} h'{}_+^{\gh_{m+1}} \ . 
\ee
Then 
\bea 
\left\{V, B^{(\tilde d)} \right\}  
      &=& 4(m{+}1)(l{+}1) \! \int \! dx \, 
d_{\am \de \gh_1 \ldots \gh_m} \tilde d_{\de \roh_1 \ldots \roh_{l+1}}
\, k_+^\am h_+^{\gh_1} \ldots h_+^{\gh_m} h_+^{\roh_1} 
                        \ldots h_+^{\roh_l} h'{}_+^{\roh_{l+1}}
\nn\\ 
&&+2(m{+}1) \! \int \! dx \, 
d_{\am \bm \gh_1 \ldots \gh_m} \tilde d_{\de \roh_1 \ldots \roh_{l+1}}
\, k_+^\am f^{\bm \de \kh} \X^\kh h_+^{\gh_1} \ldots h_+^{\gh_m} h_+^{\roh_1} 
                        \ldots h_+^{\roh_{l+1}} \qquad 
\eea
(the non-zero terms are those from brackets of $k_+$'s with $k_+$'s).
The first line can be combined with the bracket $\{ U, B^{(\tilde d)}
\}$ to produce the expression
\be
d_{a d \gh_1 \dots \gh_m} \tilde d_{b d \roh_1 \dots \roh_l}
                 h_+^{\gh_1} \ldots h_+^{\gh_m} h_+^{\roh_1} 
\ldots h_+^{\roh_l} 
       \left(k_+^a h'{}_+^b - k_+^b h'{}_+^a \right),
\ee
and this vanishes by the same argument used for (\ref{UW}).
But in addition we have 
\bea
\left\{W, B^{(\tilde d)} \right\} 
         &=& 2i \! \int \! dx \, 
d_{\deh \gh_1 \dots \gh_{m+1}} \, \tilde d_{\am \bh_1 \dots \bh_{l+1}}
\Big( 2 f_{\deh \am \kappa} k_1^{\kappa} h_+^{\gh_1} \ldots h_+^{\gh_{m+1}}
           h_+^{\bh_1} \ldots h_+^{\bh_{l+1}} \nn\\
&&  \;\;+ \ i (m{+}1)(l{+}1) \X^\deh k_+^\am f_{\gh_1 \bh_1 \kh} h_+^\kh
             h_+^{\gh_2} \ldots h_+^{\gh_{m+1}} 
h_+^{\bh_2} \ldots h_+^{\bh_{l+1}} \Big).\quad
\eea
The first of these terms is zero by invariance of $d$ and $\tilde d$, 
and the second cancels the last remaining 
term in $\{ V, B^{(\tilde d)} \}$. Hence
\be
\{ P^{(d)}, B^{(\tilde d)} \} = 0.
\ee

The final bracket we must consider is
\be \left\{F^{(e)}, P^{(d)} \right\} 
= -\left\{ F^{(e)}, U + V + W \right\}\label{OeL}.\ee
It is actually significantly easier to calculate brackets
involving the bosonic, bottom component charge 
\be
A^{(e)} = \int \! dx \, \Omega^{(e)}\left( \psi_+ ,\dots, \psi_+\right)
                     = \int \! dx \, e_{\ah_1 \dots \ah_p} 
h_+^{\ah_1} \dots h_+^{\ah_p}.
\ee
rather than the top component charge $F^{(e)}$, and the 
results are easily related by supersymmetry, as explained in section 7.1.

We find that $\{ A^{(e)} ,V\} = \{A^{(e)},W\} = 0$ and so, using
(\ref{Uh}), 
\bea
\left\{A^{(e)}, P^{(d)}\right\} &=&\left\{ A^{(e)}, U\right\}\nn\\
&=& 4p \int dx e_{\bh \ah_1 \dots \ah_{p-1}} d_{\bh \gh_1 \dots \gh_{m+1}} 
                   \left( \del_x h_+^{\ah_1} \ldots h_+^{\ah_{p-1}} 
\right) h_+^{\gh_1} \ldots h_+^{\gh_{m+1}}.
\label{edbrac}\eea
There are various ways in which this can be evaluated,
but it is convenient to note that the structure is identical to results 
obtained when investigating the current algebra of the bosonic 
PCM based on $H$ \cite{PCM} (see section 4.1 of that paper;
this deals with $H$ simple, but the conclusions 
are easily extended to direct products of groups).
We find, by comparison, that the bracket vanishes in many cases, 
for instance if $d$ and $e$ are both single trace invariants, but 
for \emph{any} choice of $d$ and $e$, the results of \cite{PCM} ensure 
that the integrand in (\ref{edbrac}) can be written as a 
polynomial in currents of type $\Omega^{(e)}$ and their $\del_+$ 
derivatives. Thus
$\{F^{(e)}, P^{(d)} \}$ does indeed have the form described in section
5.

\section{Comments}
We have presented a thorough account of two classes of local 
conservation laws in supersymmetric sigma models based on 
classical symmetric spaces $G/H$. One variety of conserved 
quantity, based on $\Omega$ forms, corresponds to the 
cohomology of the target manifold. The other variety, based on 
$\Lambda$ tensors, are higher-spin generalizations of energy-momentum.
In each case we have found primitive sets of currents, from which 
all others of $\Omega$ or $\Lambda$ type can be constructed as 
differential polynomials. The mathematics underlying the $\Omega$ type
currents is, of course, very well-understood; nevertheless, it is not 
entirely straightforward to extract the relevant sorts of results on 
primitive generators from the standard sources such as
\cite{Milnor}-\cite{Spivak:vol5}, and so the concrete presentation
of the cohomology generators which we have derived 
may, perhaps, be useful elsewhere.

We have not proved that the $\Omega$ and $\Lambda$ type currents
(and differential polynomials in them) 
give the \emph{only} local conserved quantities in these sigma models,
but all our results are consistent with this suggestion.
In particular, the Poisson bracket calculations of the top component
charges which we carried out in the last section (and whose results
were announced in section 5) always generate answers which correspond
to currents of recognizable form. 

Our Poisson bracket results generalize 
those obtained for SPCMs \cite{SPCM} and, 
as in this earlier work, we were able to 
prove the existence of commuting families of bosonic (top component)
charges. A new feature of the symmetric space models, however, is the 
appearance of fermionic top component charges, which close under
Poisson brackets onto the bosonic charges. The presence of 
commuting charges for the bosonic symmetric space model has been 
linked to the Drinfeld-Sokolov/mKdV hierarchies \cite{Evans01} and it 
would be interesting to understand how these super-generalizations fit 
into this framework.

The most obvious problem for future work is the study of these
conserved quantities at the quantum level (some recent progress
for the analogous bosonic models has recently been made in
\cite{EKMY}). Given the way in which commuting sets of bosonic 
charges are known to constrain the S-matrix, it would be very
interesting to understand what additional implications are imposed by 
the fermionic charges. This might help in the construction or 
confirmation of S-matrices, about which we still know
comparatively little for these models 
(the exceptions are the $S^n$ model \cite{Witten:77}, 
the $CP^n$ model \cite{CPn}, the $SU(n)$ super PCM \cite{EH},
and some Grassmannian theories \cite{Grass}).

Finally, there has been intense recent investigation of 
various integrable structures in string theory on 
AdS backgrounds, which are very closely related to symmetric spaces 
\cite{Mandal}-\cite{superyang}.
The models we have studied here correspond to world-sheet, rather
than target space supersymmetry in string theory. Nevertheless,
it would clearly be worthwhile to investigate and clarify how 
the structures of conserved quantities might coincide, or differ,
in these two classes of supersymmetric theories.
\vskip 10pt

{\bf Acknowledgements}: We are very grateful to Ivan Smith for helpful 
conversations concerning the cohomology of symmetric spaces.
The research of CASY was supported by a PPARC studentship; the
work of JME is supported in part by Gonville and Caius College, 
Cambridge.

\end{document}